\documentclass[aps, preprint, nofootinbib, superscriptaddress, showpacs]{revtex4-1}
\usepackage{graphicx}
\usepackage{color}
\usepackage{mathrsfs}
\usepackage[colorlinks=false]{hyperref}
\usepackage{times}

\begin{document}
\title{Quantum state tomography of molecules by ultrafast diffraction}
\author{Ming Zhang$^*$}
\affiliation{State Key Laboratory for Mesoscopic Physics and Collaborative Innovation Center of Quantum Matter, School of Physics, Peking University, Beijing 10087, China}
\author{Shuqiao Zhang}
\email{These authors contributed equally to this work.}
\affiliation{State Key Laboratory for Mesoscopic Physics and Collaborative Innovation Center of Quantum Matter, School of Physics, Peking University, Beijing 10087, China}
\author{Yanwei Xiong}
\affiliation{Department of Physics and Astronomy, University of Nebraska–Lincoln, Lincoln, NE, USA.}
\author{Hankai Zhang}
\affiliation{State Key Laboratory for Mesoscopic Physics and Collaborative Innovation Center of Quantum Matter, School of Physics, Peking University, Beijing 10087, China}
\author{Anatoly A. Ischenko}
\affiliation{Lomonosov Institute of Fine Chemical Technologies, RTU-MIREA - Russian Technological University, Vernadskii Avenue 86, 119571 Moscow, Russia}
\author{Oriol Vendrell}
\affiliation{Physikalisch-Chemisches Institut, Universit\"at Heidelberg, Im Neuenheimer Feld 229, D-69120 Heidelberg, Germany}
\author{Xiaolong Dong}
\affiliation{State Key Laboratory for Mesoscopic Physics and Collaborative Innovation Center of Quantum Matter, School of Physics, Peking University, Beijing 10087, China}
\author{Xiangxu Mu}
\affiliation{State Key Laboratory for Mesoscopic Physics and Collaborative Innovation Center of Quantum Matter, School of Physics, Peking University, Beijing 10087, China}
\author{Martin Centurion}
\affiliation{Department of Physics and Astronomy, University of Nebraska–Lincoln, Lincoln, NE, USA.}
\author{Haitan Xu}
\email{xuht@sustech.edu.cn}
\affiliation{Shenzhen Institute for Quantum Science and Engineering, Southern University of Science and Technology, Shenzhen 518055, China}
\author{R. J. Dwayne Miller}
\email{dmiller@lphys.chem.utoronto.ca}
\affiliation{Departments of Chemistry and Physics, University of Toronto, Toronto, Ontario M5S 3H6, Canada}
\author{Zheng Li}
\email{zheng.li@pku.edu.cn}
\affiliation{State Key Laboratory for Mesoscopic Physics and Collaborative Innovation Center of Quantum Matter, School of Physics, Peking University, Beijing 10087, China}

\def\ket#1{|#1\rangle}
\def\Bra#1{\langle#1|}
\def\IP#1#2{\langle#1|#2\rangle}
\def\BK#1#2#3{\langle#1|#2|#3\rangle}
\def\ketBra#1#2{|#1\rangle\langle#2|}
\def\denmat#1#2{\langle#1|\hat{\rho}|#2\rangle}
\def\Pr{\ensuremath{\mathrm{Pr}}}

\maketitle

\definecolor{oldtxtcolor}{rgb}{0.00, 0.0, 0.5}
\definecolor{newtxtcolor}{rgb}{0.00, 0.3867, 0.00}
\definecolor{newtxtcolor}{rgb}{0.00, 0.0, 1}
\definecolor{oldtxtcolor}{rgb}{1.00, 0.0, 0.00}

\def\verX{12}
\def\verO{1}
\def\verN{2}
\def\verON{12}

\ifx\verX\verO
 \newcommand { \oldtxt }[1] {{\color{oldtxtcolor}{#1}}}
 \newcommand { \newtxt }[1] {}
\fi
\ifx\verX\verN
 \newcommand { \oldtxt }[1] {}
 \newcommand { \newtxt }[1] {{\color{newtxtcolor}{#1}}}
\fi
\ifx\verX\verON
 \newcommand { \oldtxt }[1] {{\color{oldtxtcolor}{#1}}}
 \newcommand { \newtxt }[1] {{\color{newtxtcolor}{#1}}}
\fi


{\bf \noindent Ultrafast electron diffraction and time-resolved serial crystallography are the basis of the ongoing revolution in capturing at the atomic level of detail the structural dynamics of molecules. However, most experiments employ the classical “ball-and-stick” depictions, and the information of molecular quantum states, such as the density matrix, is missing. Here, we introduce a framework for the preparation and ultrafast coherent diffraction from rotational wave packets of molecules, and we establish a new variant of quantum state tomography for ultrafast electron diffraction to characterize the molecular quantum states. The ability to reconstruct the density matrix of molecules of arbitrary degrees of freedom will provide us with an unprecedentedly clear view of the quantum states of molecules, and enable the visualization of effects dictated by the quantum dynamics of molecules. 
}



\begin{figure*}[hbt!]
    \centering
    \includegraphics[width=12.0cm]{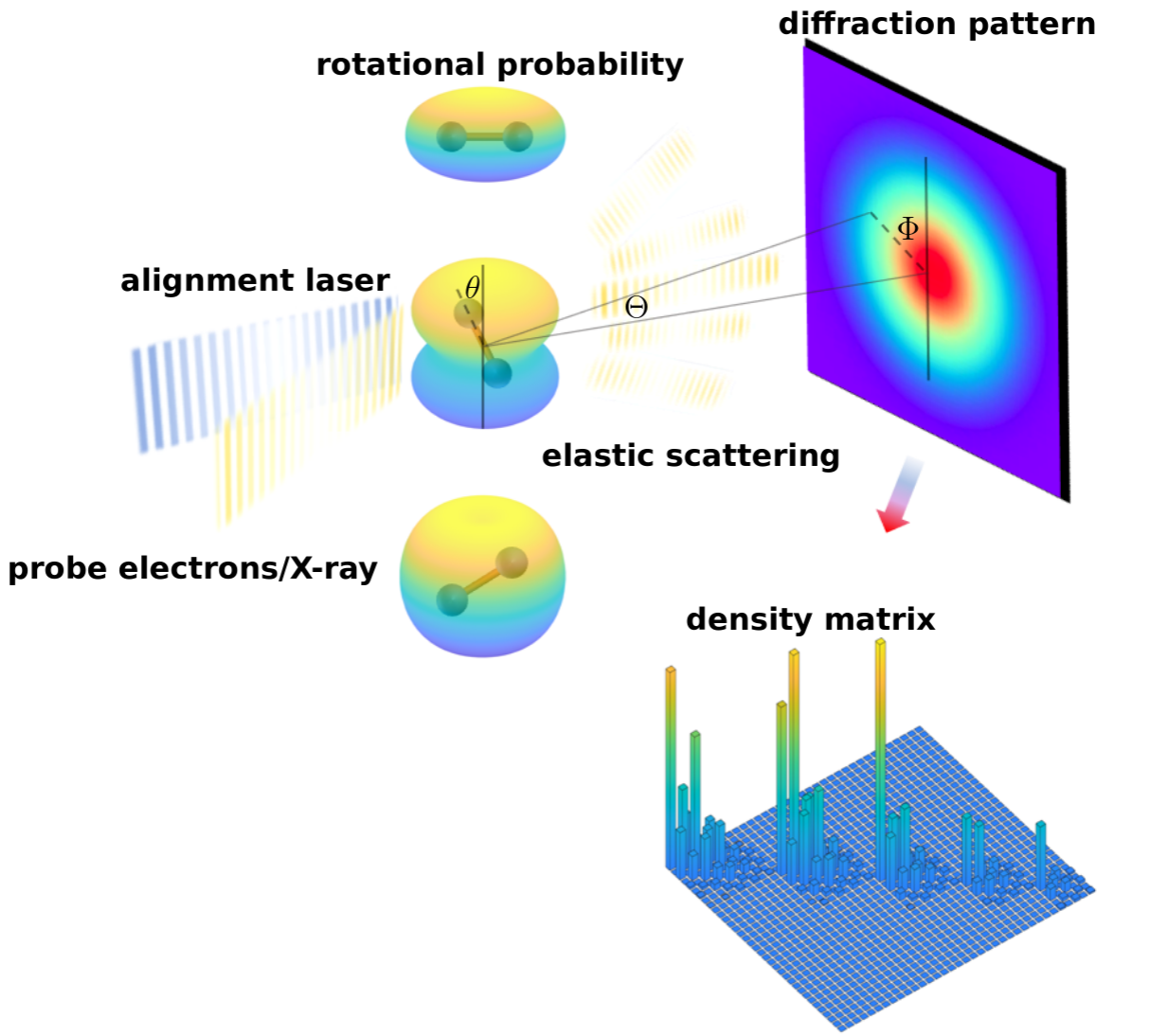}
    \caption{\label{fig:overview}
    \textbf{Schematic drawing of quantum tomography by ultrafast diffraction, illustrated with a rotational wavepacket of N$_2$ molecule.} A rotational wavepacket is prepared by an impulsive alignment laser pulse~\cite{Stapelfeldt03:RMP543}, and probed by diffraction of an incident electron/X-ray pulses for a series of time intervals. The mixed rotational quantum state represented by its density operator $\hat{\rho}$ is determined from the diffraction patterns.}
\end{figure*}

{
With the ability to directly obtain the Wigner function and density matrix of photon states, quantum tomography (QT) has made a significant impact on quantum optics~\cite{Lvovsky09:RMP299, Priebe17:NatPho11, Smithey93:PRL1244}, quantum computing~\cite{Jianwei12:Science363,Laflamme20:Nat59} and quantum information~\cite{Murch13:Nat214,saglamyurek15:83}.
By an appropriate sequence of measurements on the evolution of each degree of freedom (DOF), the full quantum state of the observed photonic system can be determined.
The first proposal to extend the application of QT to reconstruction of complete quantum states of matter wavepackets~\cite{Leonhardt96:PRL1985} had generated enormous interest in ultrafast diffraction imaging~\cite{Yangjie18:Sci64, Wolf19:NatChem504, ischenko17:CR11066,gao13:343,eichberger10:799,sciaini09:56,mehrabi19:1167,ishikawa15:1501,miller14:6175,ernstorfer09:5917,siwick03:302,wolter16:308} and pump-probe spectroscopy of molecules~\cite{Dunn95:PRL884}.
This interest was elevated with the advent of ultrafast electron and X-ray diffraction techniques using electron accelerators and X-ray free electron lasers to add temporal resolution to the observed nuclear and electron distributions~\cite{Minitti15:PRL255501,Yang16:PRL153002}.
In this respect, quantum tomography holds great promise to enable imaging of molecular wavefunctions beyond classical description.
This concept could become a natural area for quantum tomography of quantum states of matter~\cite{yang12:PRL133202,Cosmin12:Nat,YangJie16:Nat11232,Fielding18:Science30,Li19:AP296}.
However, the great interest in this area has been tempered by the illustration of an "impossibility theorem", known as the dimension problem~\cite{Mouritzen05:PRA, Mouritzen06:JCP244311}.
To obtain the density matrix of a system, the previoiusly established QT procedure relies on integral transforms (e.g. the tomographic Radon transform), which preserves dimensionality~\cite{Lvovsky09:RMP299}.
Unlike its quantum optics sibling, only a single evolutionary parameter, time, is available for the molecular wavepacket.
Not being able to associate unitary evolution to every DOF of molecular motion, quantum tomography could not be used beyond 1D and categorically excludes most vibrational and all rotational motion of molecules.

Here we present an approach to resolve the notorious dimension problem.
Solving this challenging problem is important to push imaging molecular dynamics to the quantum limit.
Our approach makes quantum tomography a truly useful method in ultrafast physics and enables the making of quantum version of a ``molecular movie"~\cite{gao13:343,miller14:6175,Fielding18:Science30, Li19:AP296,Nango16:Science1552,  Nicholson18:Science821, Weinstein12:Nat157, Yang20:Sci885}, without being limited in one dimension.
We first demonstrate this method using a numerical simulation of ultrafast diffraction imaging of laser-aligned nitrogen molecules~\cite{YangJie16:Nat11232}. The analysis with this method correctly recovers the density matrix of the rotational wavepacket (schematically shown in Fig.~\ref{fig:overview}), which is otherwise impossible to obtain with previously established QT procedures. We then apply this method to ultrafast diffraction experiments to obtain the quantum density matrix from experimental data.
}

The modern formulation of quantum tomography based on integral transform~\cite{Lvovsky09:RMP299, Leonhardt96:PRL1985, Dunn95:PRL884} originates from the retrieval of wavefunction phases lost in the measurement.
Dating back to 1933, Pauli and Feenberg proposed that a wavefunction $\psi(x,t)=|\psi(x,t)|e^{i\phi(x,t)}$ can be obtained by measuring the evolution of 1D position probability distribution $\Pr(x,t)=|\psi(x,t)|^2$ and its time derivative $\partial \Pr(x,t)/\partial t$ for a series of time points ~\cite{pauli33}. 
Equivalently, a pure quantum state can also be recovered by measuring $\Pr(x,t)$ at time $t$ and monitoring its evolution over short time intervals, i.e. $\Pr(x,t+N\Delta t)=|\psi(x,t+N\Delta t)|^2$ for $(N=0,1,2,\cdots)$.
{Reconstructing the phase of wavefunction} can be considered as the origin of quantum tomography. For a system with Hamiltonian $\hat{H}=\hat{H}_0+\hat{H}_{\mathrm{int}}$, the established 1D QT method makes use of knowledge of the non-interacting part of the Hamiltonian $\hat{H}_0$, so that its eigenfunctions can be pre-calculated and used in the tomographic reconstruction of density matrix through integral inversion transform. However, the dimension problem as demonstrated in the pioneering works~\cite{Mouritzen05:PRA, Mouritzen06:JCP244311} mathematically leads to singularity in the inversion from the evolving probability distribution to the density matrix and makes it challenging for higher dimensional QT.

We solve the QT dimension problem by exploiting the interaction Hamiltonian $\hat{H}_{\mathrm{int}}$ and the analogy between QT and crystallographic phase retrieval (CPR)~\cite{Rousse01:RMP17} in a seemingly distant field, crystallography.  
Further exploiting the interaction Hamiltonian $\hat{H}_{\mathrm{int}}$ provides us a set of physical conditions, such as the selection rules of transitions subject to $\hat{H}_{\mathrm{int}}$ and symmetry of the system. These physical conditions can be imposed as constraints in our QT approach, which is not feasible in the established QT methods based on integral transform.
By compensating with the additional physical conditions as constraints in the iterative QT procedure, the converged solution can be obtained as the admissible density matrix that complies with all the intrinsic properties of the investigated physical system.

We start by presenting the correspondence between QT and CPR. The research on CPR has been the focus of crystallography for decades~\cite{Yangjie18:Sci64,yang12:PRL133202,Yang20:Sci885,Rousse01:RMP17,Chapman11:Nat73,Seibert11:Nat78}. In crystallography, the scattered X-ray or electron wave encodes the structural information of molecules.
The measured X-ray diffraction intensity is
$I(\textbf{s})\sim |f(\textbf{s})|^2$,
where $\textbf{s}=\textbf{k}_{f}-\textbf{k}_{\mathrm{in}}$ is momentum transfer between incident and diffracted X-ray photon or electron, $f(\textbf{s})$ is the electronically elastic molecular form factor.
For X-ray diffraction, the form factor is connected to the electron density by a Fourier transform 
$f_{\mathrm{X}}(\textbf{s}) \sim \mathscr{F}[\Pr(\textbf{x})]$,
$\Pr(\textbf{x})$ is the probability density of electrons in a molecule, and $\textbf{x}$ is the electron coordinate.
The form factor of electron diffraction has a similar expression $f_{\mathrm{e}}(\textbf{s})=[\Sigma_{\alpha} N_{\alpha}\exp(i\textbf{s}\cdot\textbf{R}_{\alpha})-f_{\mathrm{X}}(\textbf{s})]/s^2$, where $N_{\alpha}$, $\textbf{R}_{\alpha}$ are the charge and position of $\alpha^{\mathrm{th}}$ nucleus.
However, the phase of the form factor, which is essential for reconstructing the molecular structure, is unknown in the diffraction experiment, only the modulus $|f(\textbf{s})|$ can be obtained from measured diffraction intensity.

\begin{figure*}[hbt!]
    \centering
    \includegraphics[width=16.0cm]{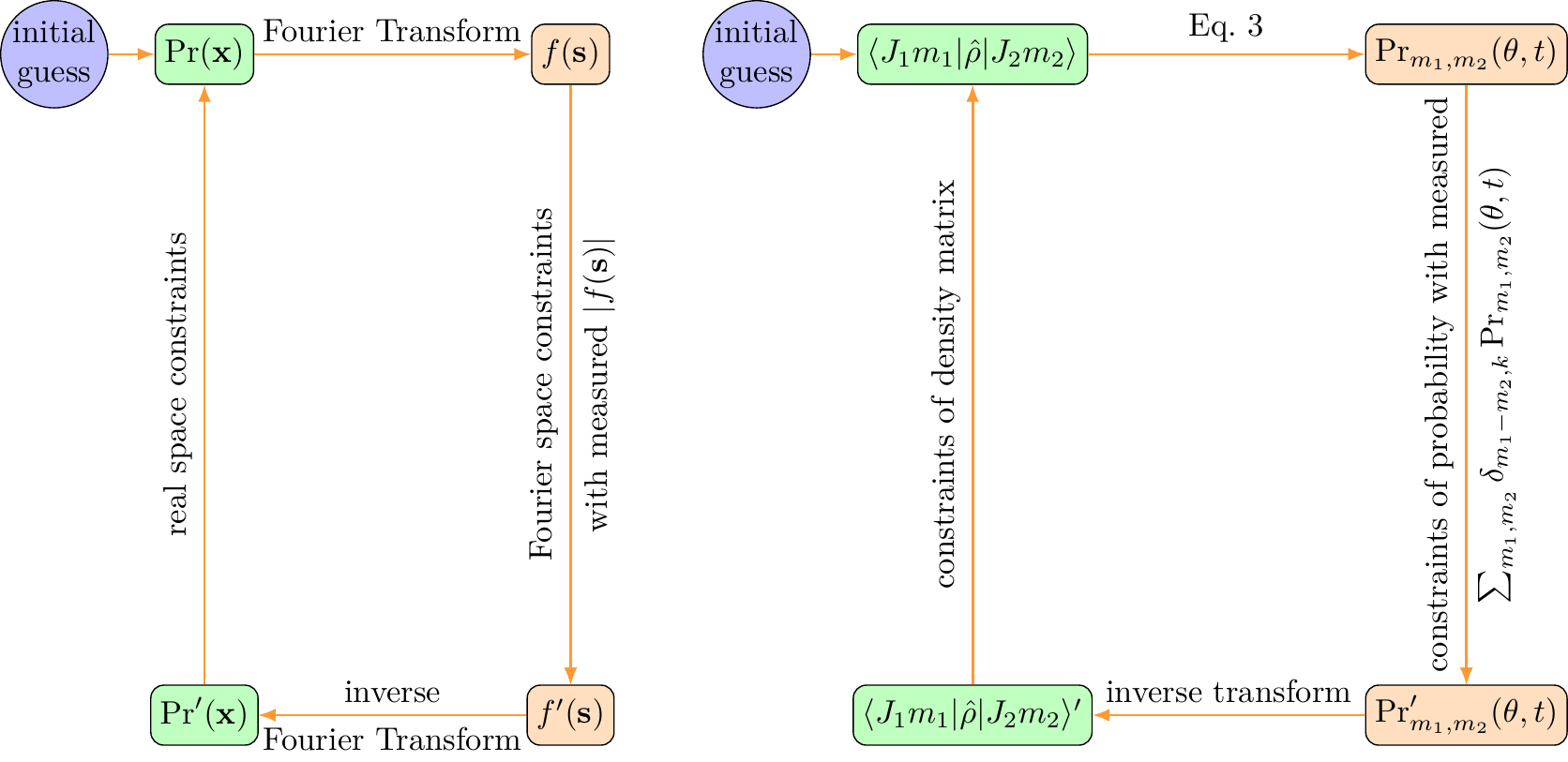}
    \caption{\label{fig1:analogy}
    \textbf{Analogy between crystallographic phase retrieval (CPR) and quantum tomography (QT) based on their common nature~\cite{pauli33}.} The CPR iterative transform between real space electron density $\Pr(\textbf{x})$ and Fourier space form factor $f(\textbf{s})$ is analogously made for QT iterative transform between blockwise probability distribution $\Pr_{m_1,m_2}(\theta,t)$ in real space and elements in density matrix space.}
\end{figure*}

Phase retrieval is a powerful method that prevails in crystallography and single particle coherent diffraction imaging~\cite{yang12:PRL133202,Chapman11:Nat73,Seibert11:Nat78}.
Its basic idea is illustrated in Fig.~\ref{fig1:analogy}. Employing projective iterations between real space and Fourier space and imposing physical constraints in both spaces, the lost phases of the form factor $f(\textbf{s})$ can be reconstructed with high fidelity. 
Fourier space constraint utilizes measured diffraction intensity data, and real space constraints comes from a priori knowledge, e.g. the positivity of electron density.
We present the new method of quantum tomography based on this conceptual approach by applying it to rotational wavepackets of nitrogen molecules prepared by impulsive laser alignment, using the ultrafast electron diffraction (UED).
Quantum tomography of rotational wavepackets is impossible in the previously established QT theory, because the full quantum state of a rotating linear molecule is a 4D object $\BK{\theta,\phi}{\hat{\rho}}{\theta',\phi'}$, while the measured probability density evolution $\mathrm{Pr}(\theta,\phi,t)$ is only 3D. It is obvious that the inversion problem to obtain the density matrix is not solvable by dimensionality-preserving transform.

From a dataset consisting of a series of time-ordered snapshots of diffraction patterns
\begin{eqnarray}
I(\textbf{s},t)=\int_0^{2\pi}d\phi\int_{0}^{\pi}\sin\theta d\theta \Pr(\theta,\phi,t) |f(\textbf{s},\theta,\phi)|^2
\,,
\end{eqnarray}
where the form factor $f$ is related to the molecule orientation. The time-dependent molecular probability distribution $\mathrm{Pr}(\theta,\phi,t)$ can be obtained by solving the Fredholm integral equation of the first kind (see supplementary information (SI) for details).
The probability distribution of a rotational wavepacket is
\begin{eqnarray}\label{Eq:den2pr}
    \Pr(\theta,\phi,t)
    = \sum_{J_1m_1}\sum_{J_2m_2}\BK{J_1m_1}{\hat{\rho}}{J_2m_2}   Y_{J_1m_1}(\theta,\phi)Y_{J_2m_2}^*(\theta,\phi)e^{-i\Delta\omega t}
    \,,
\end{eqnarray}
where $\Delta\omega=\omega_{J_1}-\omega_{J_2}$ is the energy spacing of rotational levels.
As shown in Fig.~\ref{fig1:analogy}, we devise an iterative procedure to connect the spaces of density matrix and temporal wavepacket density. 
For the system of rotational molecules, the dimension problem limits the invertible mapping between density matrix and temporal wavepacket density to the reduced density of fixed projection quantum numbers $m_1$, $m_2$,
\begin{eqnarray}\label{Eq:mblock}
\Pr_{m_1,m_2}(\theta,t)=\sum_{J_1J_2}\denmat{J_1m_1}{J_2m_2}\tilde{P}_{J_1}^{m_1}(\cos\theta)\tilde{P}_{J_2}^{m_2}(\cos\theta)e^{-i\Delta\omega t}
\,,
\end{eqnarray}
where $\tilde{P}^{m}_J(\cos\theta)$ is the normalized associated Legendre polynomial defined in SI. The analytical solution of the inverse mapping from $\Pr_{m_1,m_2}(\theta,t)$ to density matrix $\denmat{J_1m_1}{J_2m_2}$ is elaborated in SI.
However, due to the dimension problem, there is no direct way to obtain $\Pr_{m_1,m_2}(\theta,t)$ from the measured wavepacket density, only their sum is traceable through
$\sum_{m_1,m_2}\delta_{m_1-m_2,k} \Pr_{m_1,m_2}(\theta,t)
=\int_0^{2\pi} \Pr(\theta,\phi,t) e^{ik\phi} d\phi$.

Our method starts from an initial guess of density matrix and an iterative projection algorithm is used to impose constraints in the spaces of density matrix and spatial probability density.
The initial guess of quantum state, 
$\hat{\rho}_{\mathrm{ini}} = \sum_{J_0m_0}\omega_{J_0} \ketBra{J_0m_0}{J_0m_0}$, 
is assumed to be an incoherent state in the thermal equilibrium of a given rotational temperature, which can be experimentally determined~\cite{YangJie16:Nat11232}.
$\omega_{J_0} =\frac{1}{Z} g_{J_0} e^{-\beta E_{J_0}}$ is the Boltzmann weight, and $g_{J_0}$ represents the statistical weight of nuclear spin, for the bosonic $^{14}$N$_2$ molecule, $g_{J_0}$ is 6 for even $J_0$ (spin singlet and quintet) and 3 for odd $J_0$ (spin triplet).

In the probability density space, constraint is imposed by uniformly scaling each reduced density $\Pr_{m_1,m_2}(\theta,t)$ with the measured total density $\Pr(\theta,\phi,t)$. Constraints in the density matrix space enable us to add all known properties of a physical state to the QT procedure, which supply additional information to compensate the missing evolutionary dimensions.
The constraints contain general knowledge of the density matrix, i.e. the density matrix is positive semidefinite, Hermitian and with a unity trace.
Besides, the selection rules of the alignment laser-molecule interaction imply further constraints on physically nonzero $m$-blocks of the density matrix and invariant partial traces of density matrix elements subject to projection quantum number $m$ (see SI for details of the algorithm).

\begin{figure*}[hbt!]
    \centering
    \includegraphics[width=16.0cm]{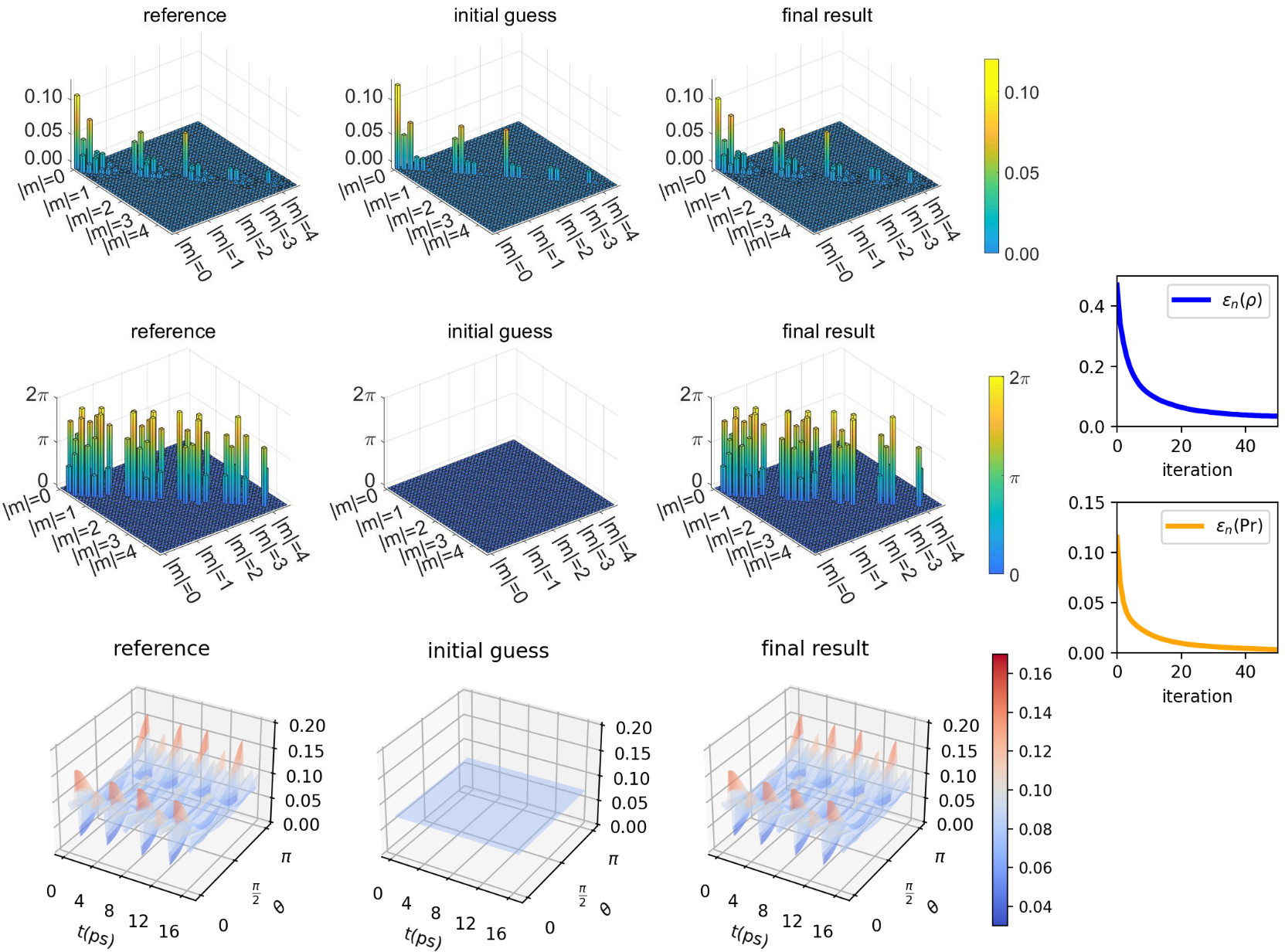}
    \caption{\label{fig:qt_simu}%
    {\bf Quantum tomography of rotational wavepacket of nitrogen molecule}. 
    The modulus and phases of density matrix elements are shown in the upper and middle panel.  {Within each $m$-block } $J=|m|,|m|+1,\cdots,J_{\mathrm{max}}$ (phases are at $t=0$). The density matrix elements of opposite magnetic quantum numbers $m$ and $-m$ are identical (see SI). Density matrix elements of higher $m$-blocks are not plotted due to their small modulus.
    The lower panel shows the wavepacket probability distribution $\Pr(\theta,t)$, which is cylindrically symmetric in azimuthal direction of $\phi$.
    The convergence of the procedure is illustrated in the rightmost column.}
\end{figure*}

We first demonstrate the capability of our approach to correctly recover the density matrix despite the dimension problem, using numerical simulation of ultrafast diffraction of impulsively aligned nitrogen molecule with an arbitrarily chosen temperature of 30 K.
The order of recovered density matrix sets the requirement on the resolution. 
From Eq.~\ref{Eq:mblock}, the characteristic time scale of rotation is $\frac{1}{\Delta\omega}=\frac{2\mathcal{I}}{|\Delta J|(J+1)}$, where $\mathcal{I}$ is the moment of inertia of nitrogen molecule, $\Delta J=J_1-J_2$ and $J=J_1+J_2$ for any two eigenstates with $J_1, J_2$.
Using the Nyquist–Shannon sampling theorem, the required temporal resolution $\delta t$ should be  
$\delta t\leq\frac{1}{2\Delta\omega}$.
The spatial resolution $\delta \theta$ and $\delta \phi$ can be determined with the argument that the nodal structure of spherical harmonic basis in Eq.~\ref{Eq:den2pr} must be resolved, i.e. 
$\delta \theta<\frac{\pi}{2J_{\mathrm{max}}}$.
To recover density matrix up to the order $J_{\mathrm{max}}=8$, it demands time resolution $\delta t \sim 10^{2}$ fs and spatial resolution $\delta \theta \sim 10^{-1}$ rad. 
Quantum tomography of the rotational wavepacket gives the result shown in Fig.~\ref{fig:qt_simu}. After 50 iterations, both density matrix and probability distribution are precisely recovered. The error of density matrix is $\epsilon_{50}(\hat{\rho})=2.9\times 10^{-2}$ and error of probability achieves $\epsilon_{50}(\Pr)=3.8\times 10^{-5}$ (see SI for the definition of $\epsilon(\hat{\rho})$ and $\epsilon(\Pr)$).

\begin{figure*}[hbt!]
    \centering
    \includegraphics[width=16.0cm]{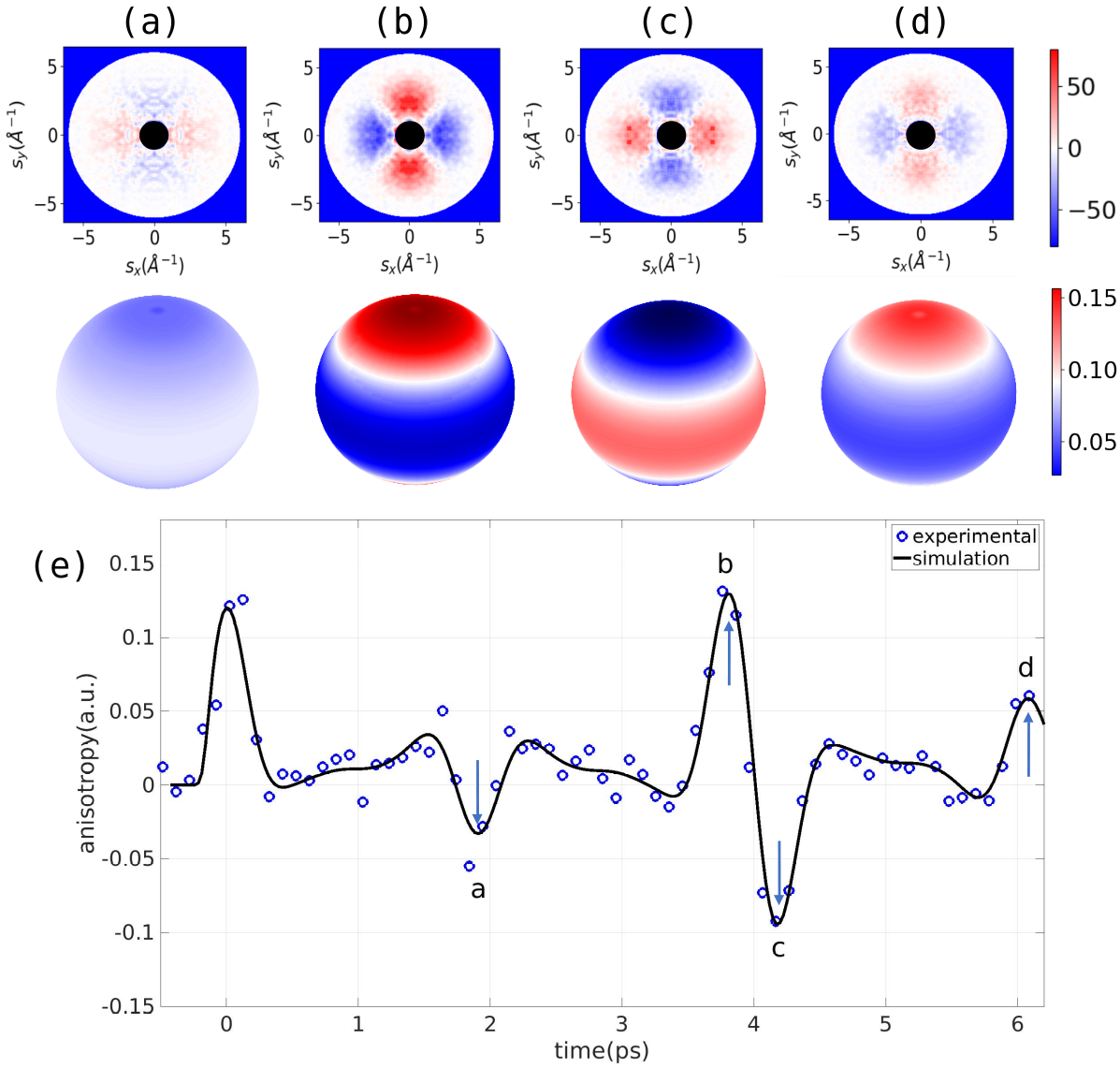}
    \caption{\label{fig:diffpic_experiment}
    {\textbf{Experimental UED data for N$_2$ rotational wavepacket.}}
    Difference-diffraction pattern and the angular probability distribution $\Pr(\theta,\phi,t)$ at various delay times marked in (e): (a) $t = 1.9$ ps, (b) $t = 3.8$ ps,  (c) $t = 4.2$ ps, (d) $t = 6.1$ ps. The dark circle corresponds to the regions where scattered electrons are blocked by the beam stop. 
    (e) Temporal evolution of the experimental and simulated anisotropy of the rotational wavepacket. 
    }
\end{figure*}

\begin{figure*}[hbt!]
    \centering
    \includegraphics[width=16.0cm]{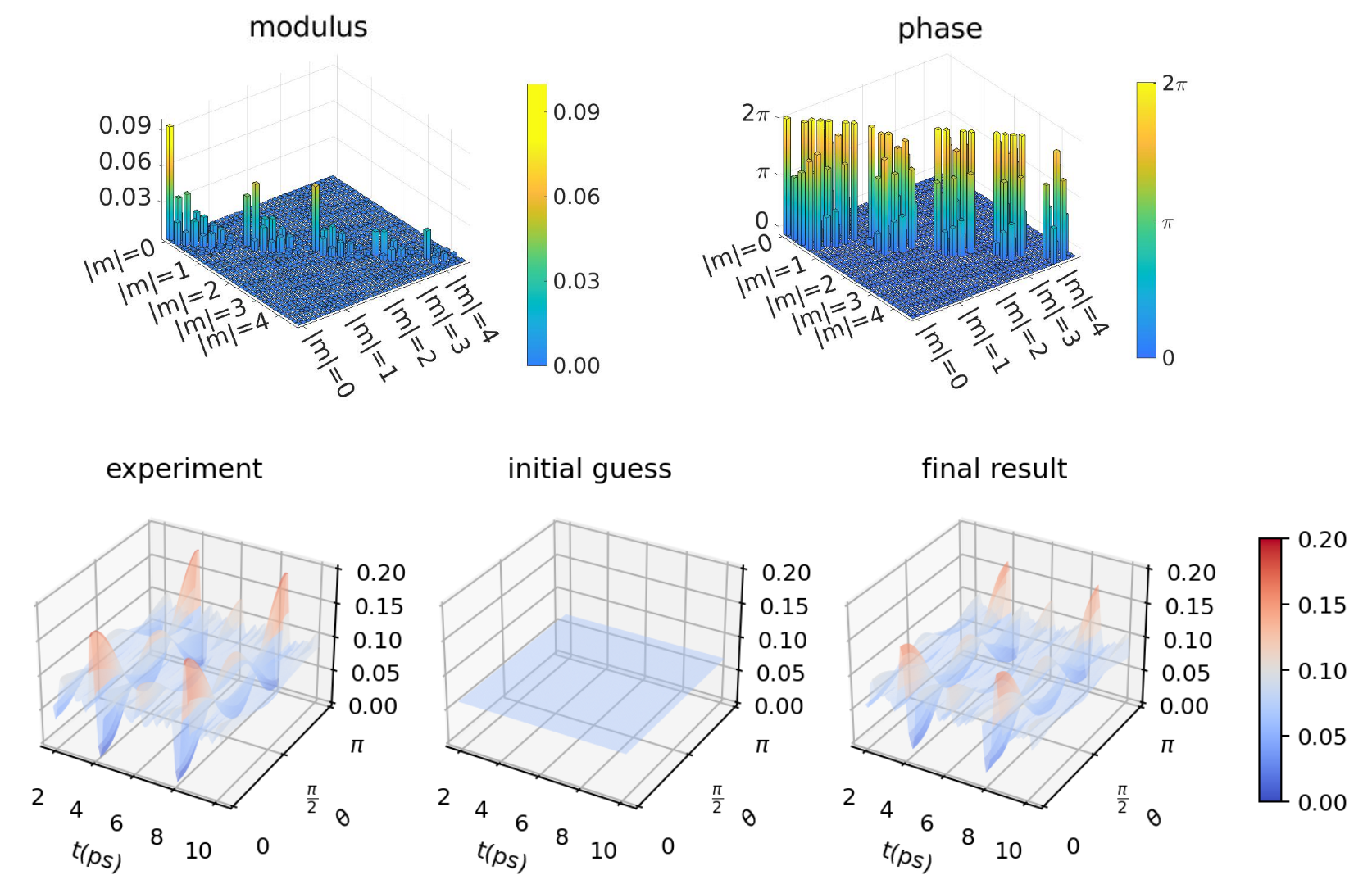}
    \caption{\label{fig:qt_exp}%
    \textbf{Experimental quantum tomography of rotational wavepacket of nitrogen molecule}. 
    The modulus and phases of QT retrieved density matrix elements are shown in the upper panel. Within each $m$-block $J=|m|,|m|+1,\cdots,J_{\mathrm{max}}$ (phases are plotted at $t=1.95$~ps after the alignment pulse). The density matrix elements of opposite magnetic quantum numbers $m$ and $-m$ are identical (see SI). Density matrix elements of higher $m$-blocks are not plotted due to their small modulus.
    The lower panel shows the wavepacket probability distribution $\Pr(\theta,t)$ (cylindrically symmetric in azimuthal direction of $\phi$) of experimental data, initial guess and final result of QT.}
\end{figure*}

We then apply this iterative QT method to the ultrafast electron diffraction (UED) experiment to extract the quantum density matrix of N$_2$ rotational wavepacket, prepared at a temperature of 45 K.
{ {The experimental parameters} are described in detail in a previous publication \cite{ Xiong:2020}. We use a tabletop kilo-electron-volt (keV) gas-phase UED setup to record the diffraction patterns of nitrogen molecules that are impulsively aligned by a femtosecond laser pulse. The details of the keV UED setup has been introduced in \cite{ Xiong:2020, Zandi:2017}, which is schematically shown in Fig.~\ref{fig:overview}. Briefly, an 800 nm pump laser pulse with a pulse duration of 60 fs (FWHM) and pulse energy of 1 mJ is used to align the molecules. A probe electron pulse with kinetic energy of 90 keV and 10,000 electrons per pulse  {is used and the diffraction pattern of the electrons scattered from the molecules is recorded.} The nitrogen molecules are introduced in a gas jet using a de Laval nozzle. The laser pulse has a tilted pulse front to compensate the group velocity mismatch between the laser and electron pulses, and an optical stage is used to control the time delay between the pump and probe pulse with a time step of 100 fs. The pump laser launches a rotational wave packet, which exhibits dephasing and subsequent revivals of alignment in picosecond time scale. The experimental diffraction patterns at  {several} time delays are shown in Fig.~\ref{fig:diffpic_experiment}(a)-(d). The temporal evolution of diffraction patterns can be characterized by the anisotropy, defined as $(S_H-S_V)/(S_H+S_V)$, where $S_H$ and $S_V$ are the sum of the counts in horizontal and vertical cones in the diffraction patterns at $3.0<s<4.5$ \AA$^{-1}$, with an opening angle of 60 degrees.  {The temporal evolution of angular probability distribution $\Pr(\theta,\phi,t)$ can be retrieved} using the method described in \cite{Xiong:2020}, followed by a deconvolution using a point spread function with FWHM width of 280 fs to remove the blurring effect due to the limited temporal resolution of the setup. Data is recorded from before excitation of the laser up to 6.1 ps after excitation. In order to complete the data up to a full cycle, which is needed for the quantum tomography, the angular probability distribution evolution is extended to obtain the data from 6.1 ps to 11 ps using a reflection of the data from 6.1 ps to 1.2 ps based on the symmetry of the evolution of the rotational wavepacket. The diffraction patterns and corresponding angular distributions at  {various} time delays are shown in Fig.~\ref{fig:diffpic_experiment}.}
Using our QT method, we obtain the complex density matrix in Fig.~\ref{fig:qt_exp}, which completely determines the rotational quantum state of the system. The error of recovered probability distribution converges to $\epsilon(\Pr)=6.4\times 10^{-2}$.
 {The difference between recovered angular probability distribution and the experimental result comes from the restriction of order of recovered density matrix due to limited temporal and angular resolution in the experiment.}

In summary, we have demonstrated an iterative quantum tomography approach that is capable of extracting the density matrix of high-dimensional wavepacket of molecules from its evolutionary probability distribution in time.
The notorious dimension problem, which has prohibited for almost two decades the quantum tomographic reconstruction of molecular quantum state from ultrafast diffraction, has thus been resolved. This quantum tomography approach can be straightforwardly extended to obtain quantum states of vibrational wavepackets and electronic degrees of freedom as well (see SI).
We expect this advance to have a broad impact in many areas of science and technology, not only for making the quantum version of molecular movies, but also for QT of other systems when quantum state information is tainted by insufﬁcient evolutionary dimensions or incomplete measurements.

\bibliographystyle{naturemag}
\bibliography{QT}

\section*{Data availability}
The data that support the plots within this paper
and other findings of this study are available from the corresponding
authors upon reasonable request.

\section*{Acknowledgements}
We thank Jie Yang, Yi-Jen Chen, Zunqi Li and Stefan Pabst for useful discussions. This work was supported by NSFC Grant No. 11974031, funding from state key laboratory of mesoscopic physics, RFBR Grant No. 20-02-00146, and NSERC. Y.X. and M.C. were supported by the National Science Foundation, Physics Division, Atomic, Molecular and Optical Sciences program, under Award No. PHY-1606619.

\section*{Author Contribution}
Z.L. designed the study. M.Z., S.Q.Z., H.K.Z. and Z.L. carried out the calculations. M.Z., S.Q.Z., H.K.Z., X.L.D., X.X.M, H.X., O.V., R.J.D.M., A.I. and Z.L. analysed the data. Y.X., M.C., and M.Z. analyzed the experimental data. All authors contributed to the writing of the manuscript.

\section*{Competing ﬁnancial interests}
The authors declare no competing financial interests.

\clearpage

\end{document}



\title{{Supplementary Information for}\\ Quantum state tomography of molecules by ultrafast diffraction}

\author{Ming Zhang$^*$}
\affiliation{State Key Laboratory for Mesoscopic Physics and Collaborative Innovation Center of Quantum Matter, School of Physics, Peking University, Beijing 10087, China}
\author{Shuqiao Zhang$^*$}
\email{These authors contributed equally to this work.}
\affiliation{State Key Laboratory for Mesoscopic Physics and Collaborative Innovation Center of Quantum Matter, School of Physics, Peking University, Beijing 10087, China}
\author{Yanwei Xiong}
\affiliation{Department of Physics and Astronomy, University of Nebraska–Lincoln, Lincoln, NE, USA.}
\author{Hankai Zhang}
\affiliation{State Key Laboratory for Mesoscopic Physics and Collaborative Innovation Center of Quantum Matter, School of Physics, Peking University, Beijing 10087, China}
\author{Anatoly A. Ischenko}
\affiliation{Lomonosov Institute of Fine Chemical Technologies, RTU-MIREA - Russian Technological University, Vernadskii Avenue 86, 119571 Moscow, Russia}
\author{Oriol Vendrell}
\affiliation{Physikalisch-Chemisches Institut, Universit\"at Heidelberg, Im Neuenheimer Feld 229, D-69120 Heidelberg, Germany}
\author{Xiaolong Dong}
\affiliation{State Key Laboratory for Mesoscopic Physics and Collaborative Innovation Center of Quantum Matter, School of Physics, Peking University, Beijing 10087, China}
\author{Xiangxu Mu}
\affiliation{State Key Laboratory for Mesoscopic Physics and Collaborative Innovation Center of Quantum Matter, School of Physics, Peking University, Beijing 10087, China}
\author{Martin Centurion}
\affiliation{Department of Physics and Astronomy, University of Nebraska–Lincoln, Lincoln, NE, USA.}
\author{Haitan Xu}
\email{xuht@sustech.edu.cn}
\affiliation{Shenzhen Institute for Quantum Science and Engineering, Southern University of Science and Technology, Shenzhen 518055, China}
\author{R. J. Dwayne Miller}
\email{dmiller@lphys.chem.utoronto.ca}
\affiliation{Departments of Chemistry and Physics, University of Toronto, Toronto, Ontario M5S 3H6, Canada}
\author{Zheng Li}
\email{zheng.li@pku.edu.cn}
\affiliation{State Key Laboratory for Mesoscopic Physics and Collaborative Innovation Center of Quantum Matter, School of Physics, Peking University, Beijing 10087, China}

\def\ket#1{|#1\rangle}
\def\Bra#1{\langle#1|}
\def\IP#1#2{\langle#1|#2\rangle}
\def\BK#1#2#3{\langle#1|#2|#3\rangle}
\def\ketBra#1#2{|#1\rangle\langle#2|}
\def\denmat#1#2{\langle#1|\hat{\rho}|#2\rangle}
\def\dist#1#2{ \Vert #1-#2 \Vert }
\def\Pr{\ensuremath{\mathrm{Pr}}}
\def\Tr{\ensuremath{\mathrm{Tr}}}
\def\Re{\ensuremath{\mathrm{Re}}}
\def\Im{\ensuremath{\mathrm{Im}}}
\def\odd{\ensuremath{\mathrm{odd}}}
\def\even{\ensuremath{\mathrm{even}}}


\def\ket#1{|#1\rangle}
\def\Bra#1{\langle#1|}
\def\IP#1#2{\langle#1|#2\rangle}
\def\BK#1#2#3{\langle#1|#2|#3\rangle}
\def\ketBra#1#2{|#1\rangle\langle#2|}
\def\denmat#1#2{\langle#1|\hat{\rho}|#2\rangle}


\definecolor{oldtxtcolor}{rgb}{0.00, 0.0, 0.5}
\definecolor{newtxtcolor}{rgb}{0.00, 0.3867, 0.00}
\definecolor{newtxtcolor}{rgb}{0.00, 0.0, 1}
\definecolor{oldtxtcolor}{rgb}{1.00, 0.0, 0.00}

\def\verX{12}
\def\verO{1}
\def\verN{2}
\def\verON{12}

\ifx\verX\verO
 \newcommand { \oldtxt }[1] {{\color{oldtxtcolor}{#1}}}
 \newcommand { \newtxt }[1] {}
\fi
\ifx\verX\verN
 \newcommand { \oldtxt }[1] {}
 \newcommand { \newtxt }[1] {{\color{newtxtcolor}{#1}}}
\fi
\ifx\verX\verON
 \newcommand { \oldtxt }[1] {{\color{oldtxtcolor}{#1}}}
 \newcommand { \newtxt }[1] {{\color{newtxtcolor}{#1}}}
\fi

\maketitle

\clearpage
\begin{figure}[hbt!]
    \centering
    \includegraphics[width=12.0cm]{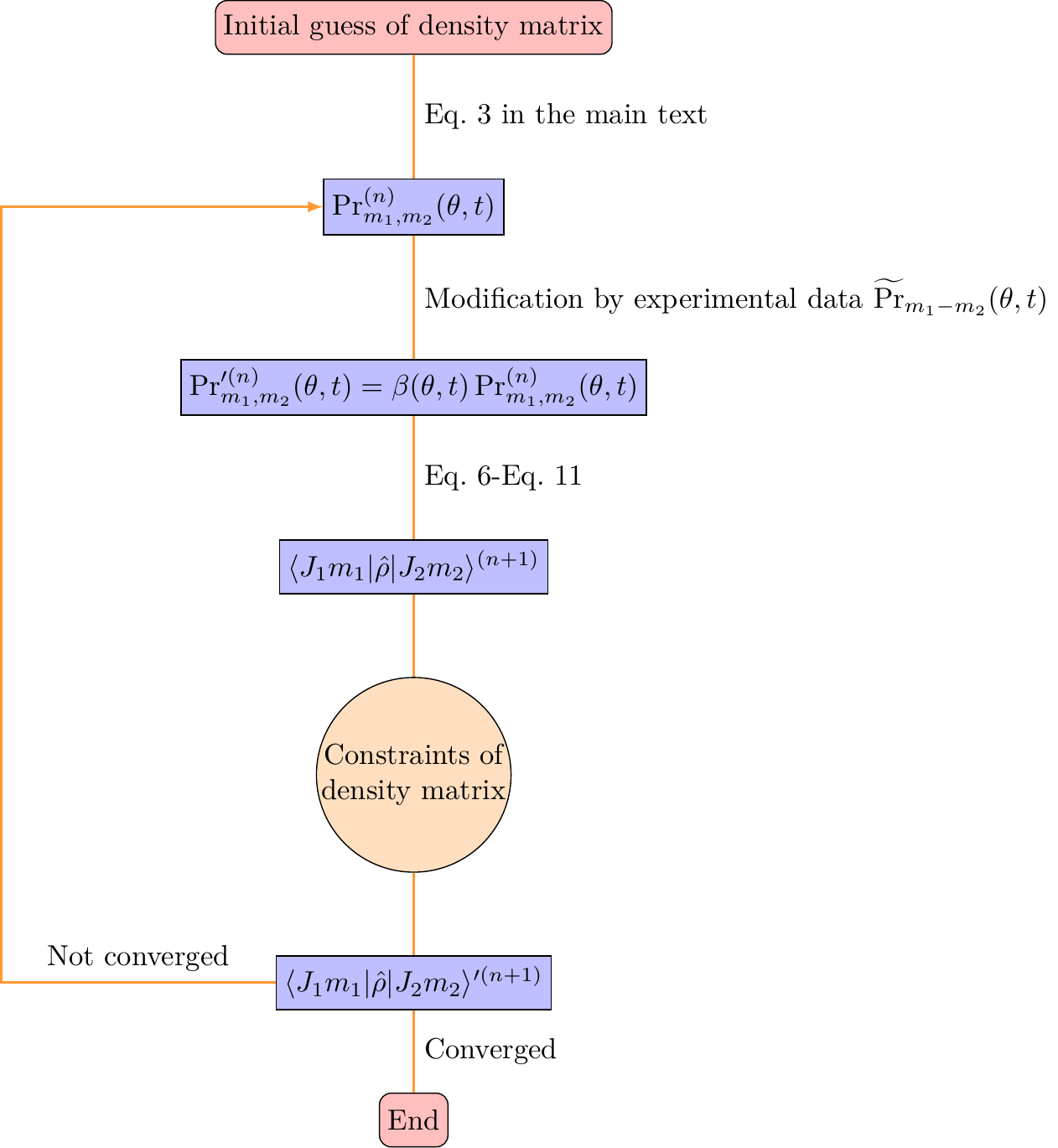}
    \caption{\label{fig:constraint_Pr}%
    {\bf Schematic flow chart for imposing constraints to the wavepacket probability distribution}. The internal procedure for the "constraints of density matrix" is separately elaborated in Fig.~\ref{fig:constraint_dm}. The superscript $n$ represents $n$-th iteration.
    }
\end{figure}
\begin{figure}[hbt!]
    \centering
    \includegraphics[width=10.0cm]{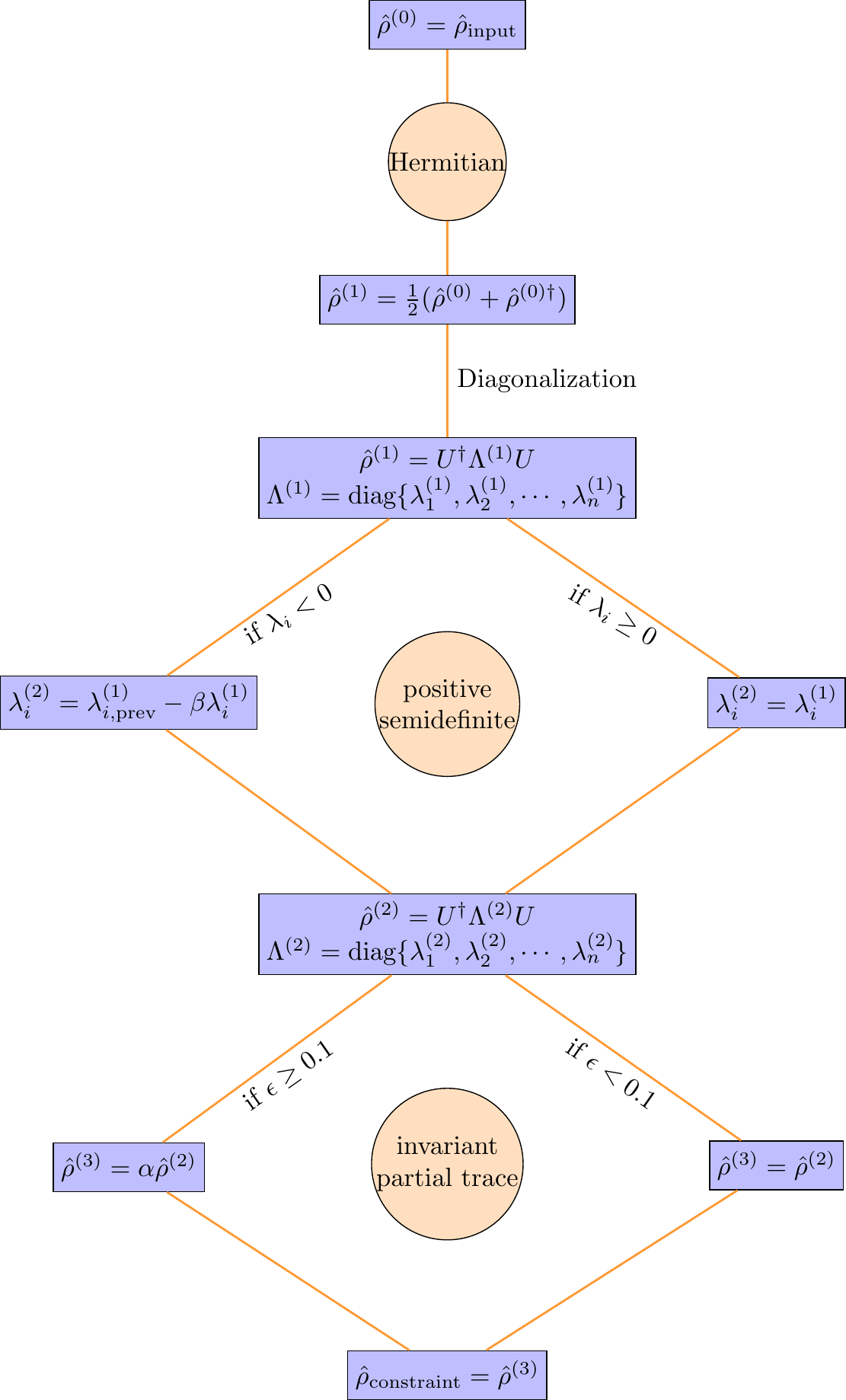}
    \caption{\label{fig:constraint_dm}%
    {\bf Schematic flow chart for imposing constraints to the density matrix}. Here $\epsilon=\left|\frac{\sum_{J_{\odd}}-\sum_{J_{0\,\odd}}\omega_{J_0}}{\sum_{J_{0\,\odd}}\omega_{J_0}}\right|$. $\alpha$ is defined in Eq.~\ref{Eq:alp}.
    We use hybrid input-output (HIO) algorithm for the positivity constraint with $\beta=0.9$~\cite{Pabst10:PRA043425}, where the subscript "prev" stands for the use of values in the previous iteration.
    }
\end{figure}
\begin{figure*}[hbt!]
    \centering
    \includegraphics[width=11.0cm]{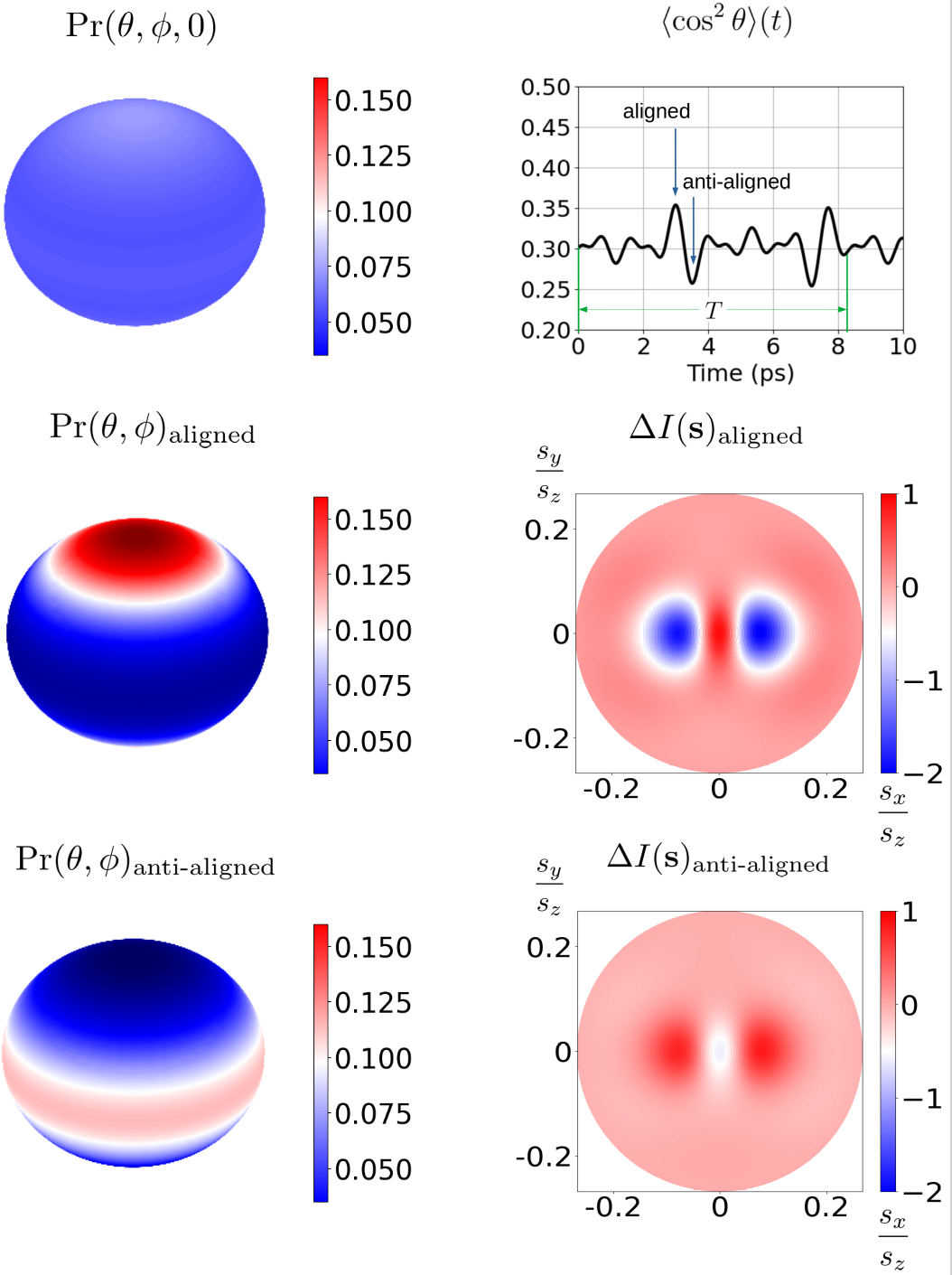}
    \caption{\label{fig:diffpic_sim}%
    {\bf Simulated probability distribution and diffraction pattern of rotational wavepacket}.
    The first row shows the initial angular probability for N$_2$ molecules prepared at a rotational temperature of 30 K and the expectation values of
     $\cos^2\theta$ of the time evolving wavepacket for N$_2$ molecules after laser pulse~\cite{Stapelfeldt03:RMP543}.
     The alignment laser pulse is linearly polarized with a Gaussian envelope of duration $\tau_L=50$ fs and $10^{13}$ W/cm$^2$ peak intensity, and $\theta$ is the polar angle between the polarization and the molecular axes. The duration is much shorter than the characteristic rotational time $\tau_L\ll T$.
     The second and third rows show the angular probability distribution changes from aligned to anti-aligned, and the difference of their diffraction intensity with respect to $t=0$. The X-ray photon energy is assumed to be 20 keV.}
\end{figure*}
\begin{figure}[hbt!]
    \centering
    \includegraphics[width=\textwidth]{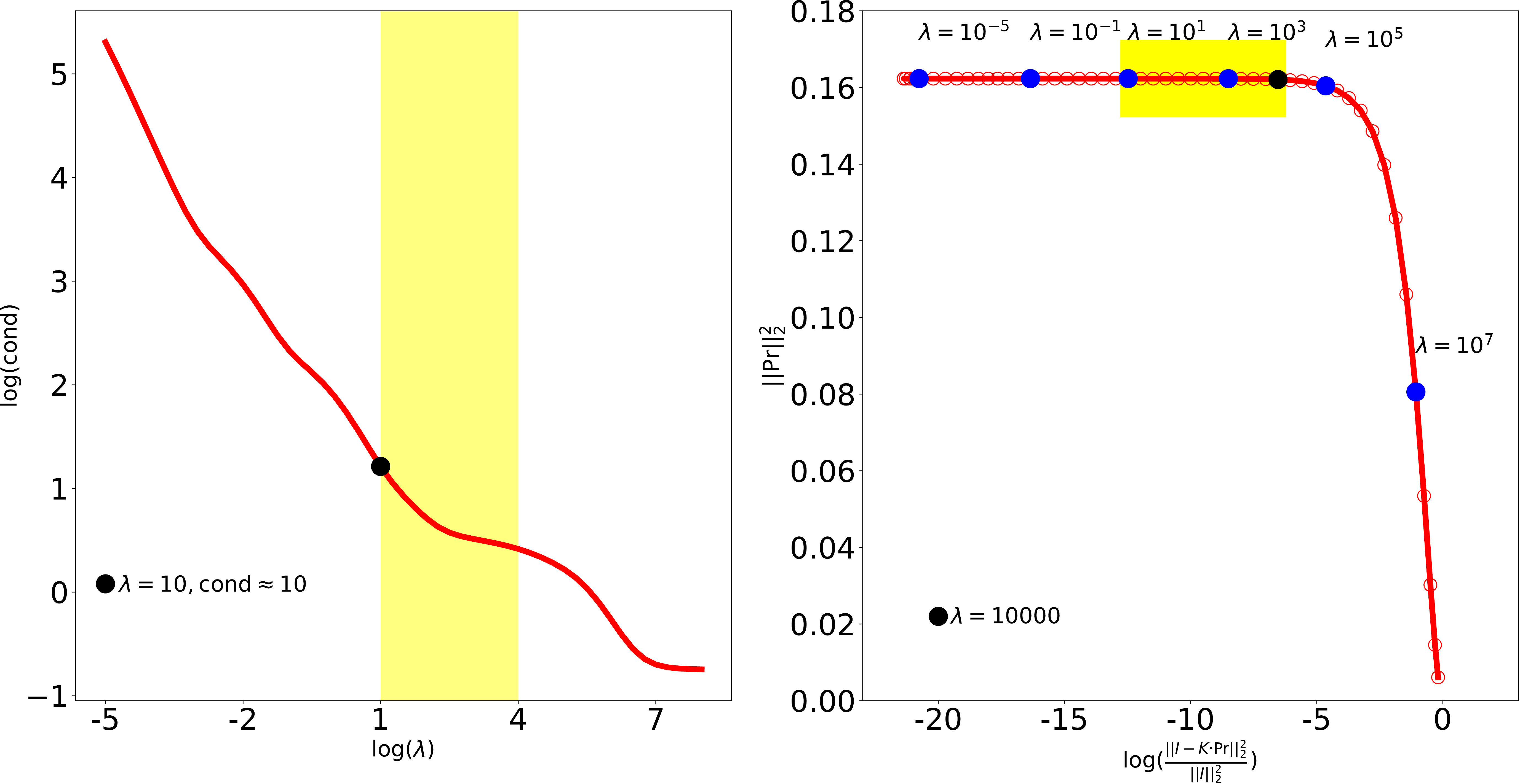}
    \caption{\label{fig:Sensitivity_Analysis}%
{\bf Faithfulness of the probability distribution Pr obtained from integral equation with Tikhonov regularization}.
(Left) Logarithm of condition number versus logarithm of the regularization parameter $\lambda$. Larger $\lambda$ makes the problem more insensitive to the measurement error $\Delta I$. 
The approximate position of the black point marked on the sketch is (1,1) (we use an approximate position because every calculation that contains generation of the random numbers leads to slightly different curve). 
(Right) The values of $\|\mathrm{Pr}\|_{2}^{2}$ and the residual $\log(\frac{\|I-\mathbf{K}\cdot\mathrm{Pr}\|_{2}^{2}}{\|I\|_{2}^{2}})$ for $\lambda$ ranging from $10^{-5}$ to $10^{8}$. 
The Tikhonov regularization procedure minimizes $\|I-\mathbf{K}\mathrm{Pr}\|_{2}^{2}+\lambda\|\mathrm{Pr}\|_{2}^{2}$. 
The black point marked on the curve is the turning point corresponding to $\lambda\approx{10^{4}}$.
The yellow area starting from $\log{\lambda}=1$ and ending at $\log{\lambda}=4$ illustrates the admissible range of regularization parameter $\lambda$.
    }
\end{figure}
\begin{figure}[hbt!]
    \centering
    \includegraphics[width=16.0cm]{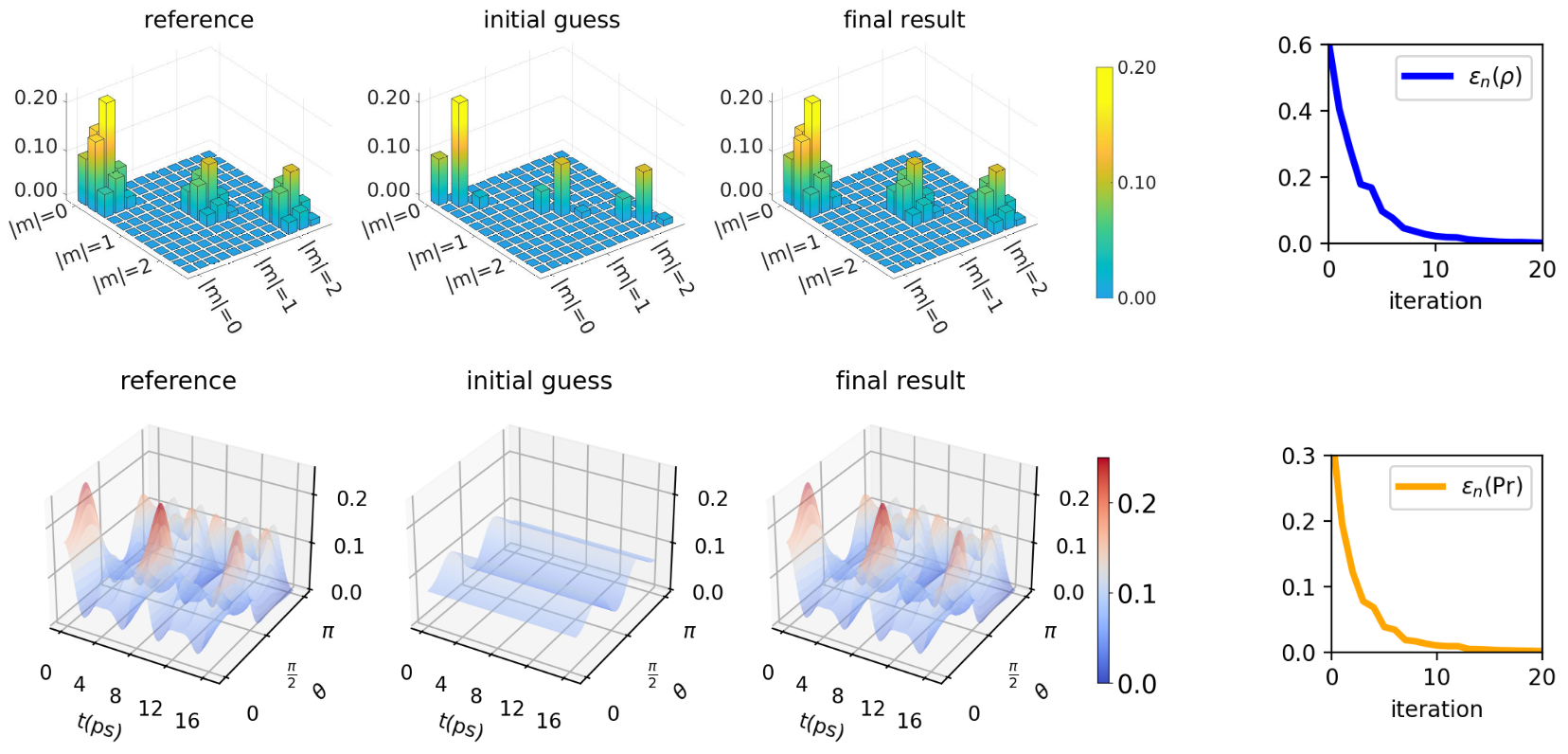}
    \caption{\label{Fig:NumTrial}
    {\bf Quantum tomography result of numerical trial with initial guess of  {correct diagonal elements of density matrix}}. The modulus of density matrix elements are shown in the upper panel, where $J=|m|,|m|+1,\cdots,J_{\mathrm{max}}$ within each $m$-block.  {The phases of all density matrix elements are zero at $t=0$.} The lower panel shows angular probability distribution, the recovered modulus and phases of density matrix elements faithfully reproduce the reference $\Pr(\theta,t)$, which is cylindrically symmetric in azimuthal direction of $\phi$. Error functions of density matrix and probability distribution are shown in the rightmost column.}
\end{figure}
\begin{figure}[hbt!]
    \centering
    \includegraphics[width=16.0cm]{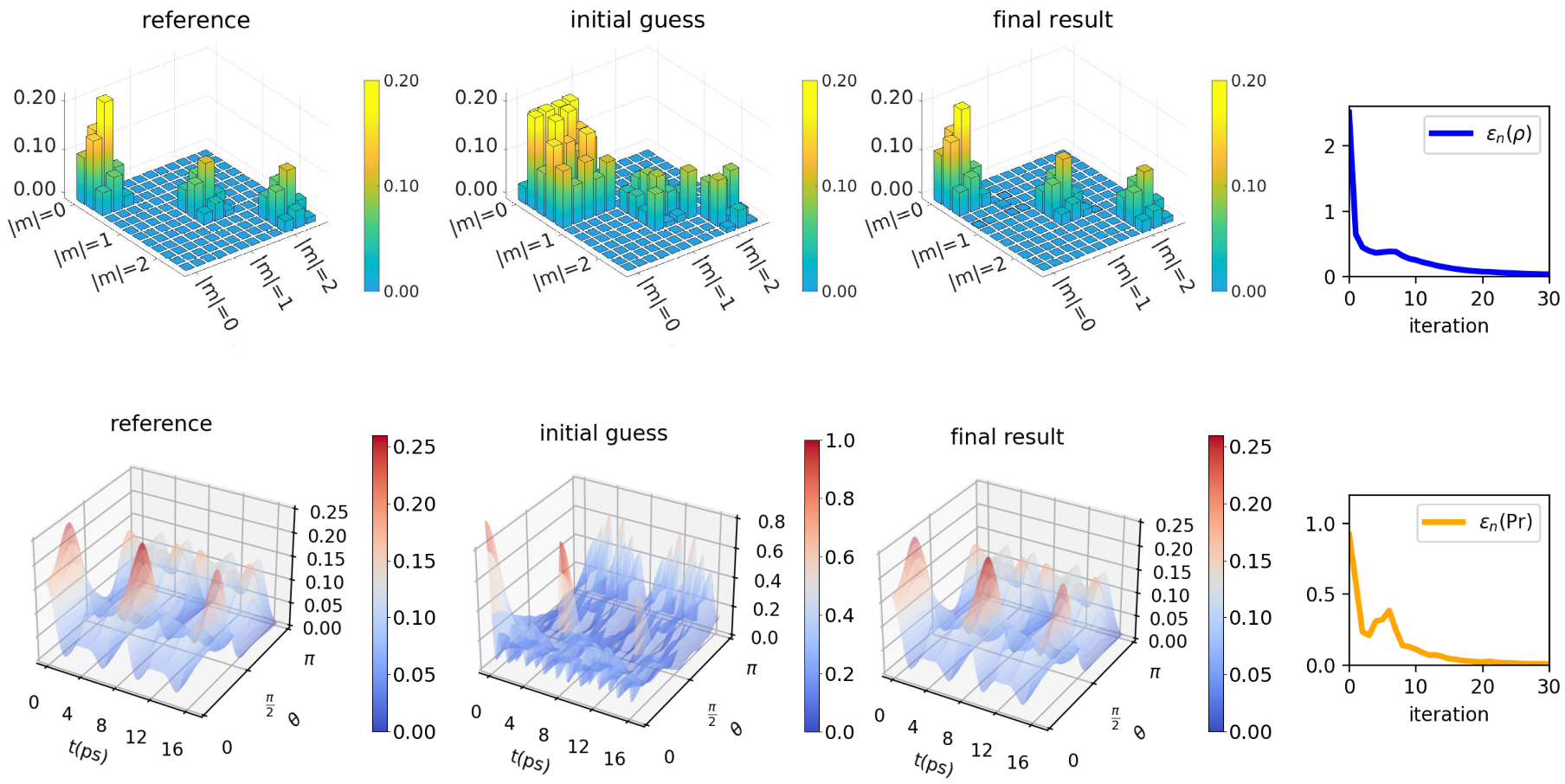}
    \caption{\label{Fig:rdm}
    {\bf Quantum tomography result of numerical trial with random initial guess of density matrix}. Only the measured probability distribution and general properties of density matrix (being Hermitian, positive semidefinite and with unity trace) are imposed as constraints during the iteration algorithm. The density matrix to be recovered and its probability distribution are identical to that in Fig.~\ref{Fig:NumTrial}.
    The modulus of density matrix elements are shown in the upper panel, where $J=|m|,|m|+1,\cdots,J_{\mathrm{max}}$ within each $m$-block.  {The phases of all density matrix elements are zero at $t=0$.} The lower panel shows angular probability distribution, the recovered modulus and phases of density matrix elements faithfully reproduce the reference $\Pr(\theta,t)$ (cylindrically symmetric in azimuthal direction of $\phi$). Error functions of density matrix and probability distribution are shown in the rightmost column.
    }
\end{figure}
\begin{figure*}[hbt!]
    \centering
    \includegraphics[width=12.0cm]{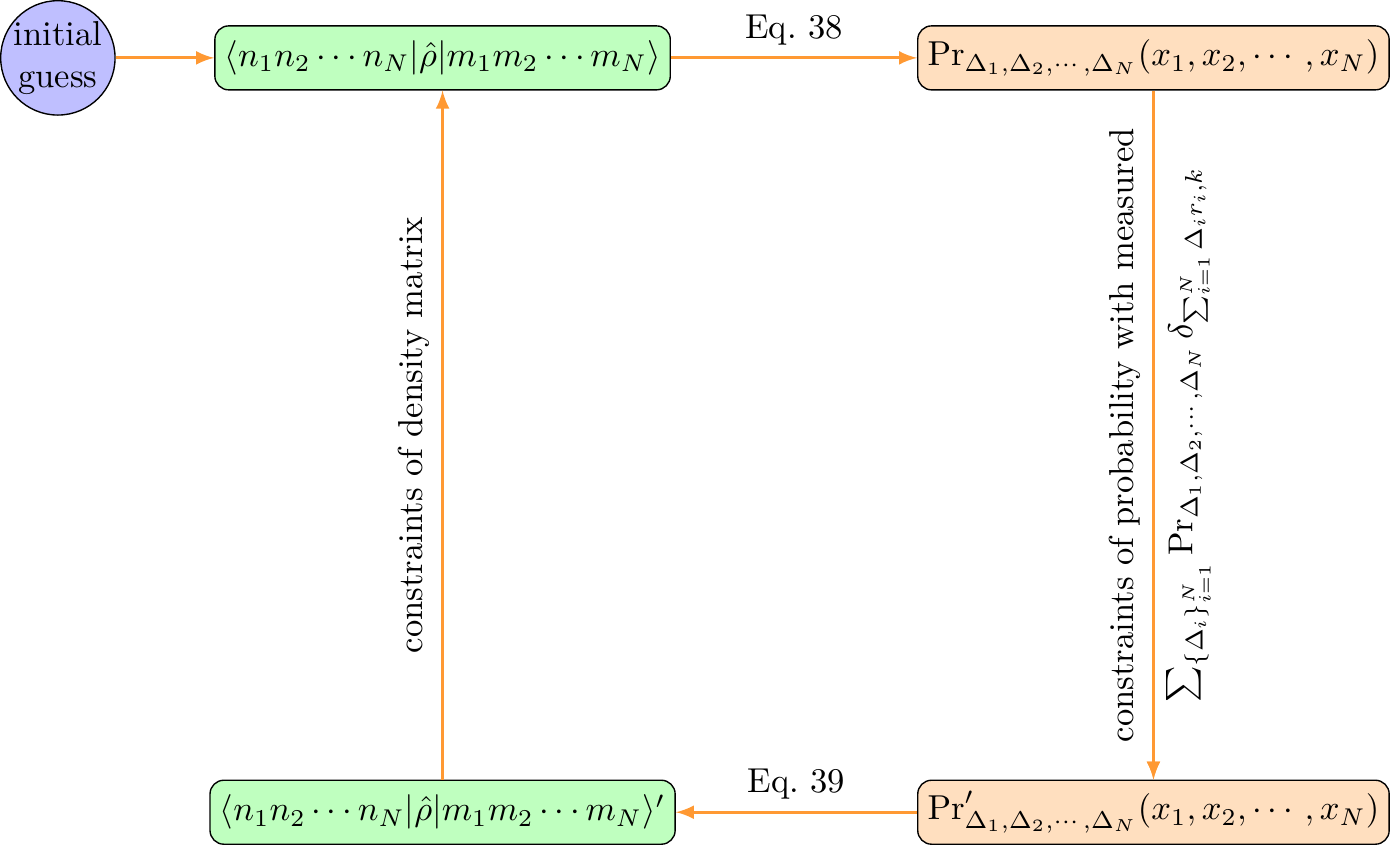}
    \caption{\label{fig:vib}%
    \textbf{Quantum tomography of vibrational state.}
    The iterative transform is again between the spaces of density matrix and the blockwise probability distribution $\Pr_{\Delta_1,\Delta_2,\cdots,\Delta_N}(x_1,x_2,\cdots,x_N)$.}
\end{figure*}
\begin{figure*}[hbt!]
    \centering
    \includegraphics[width=16.0cm]{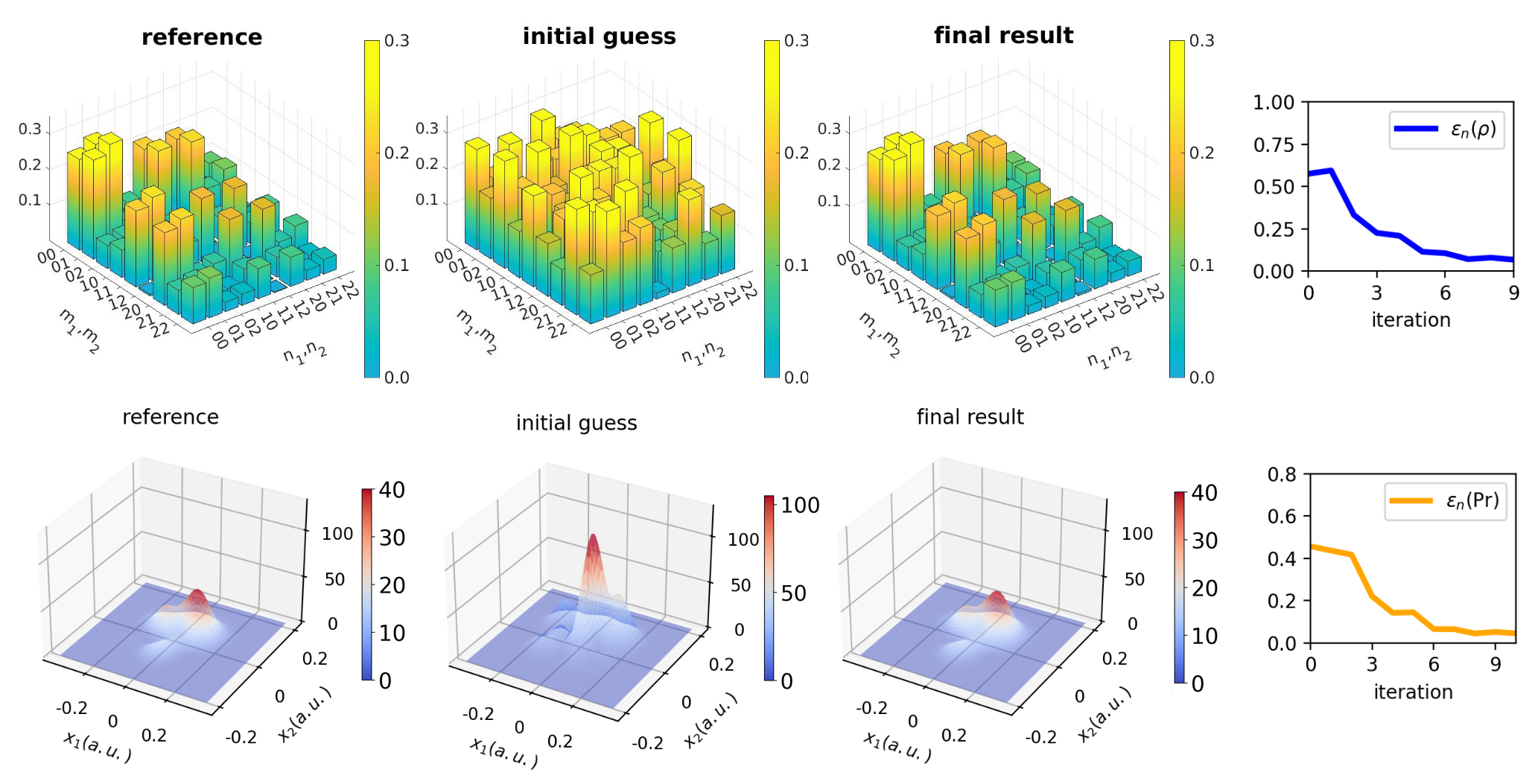}
    \caption{\label{fig:vibqtres}%
    {\bf Quantum tomography of two-dimensional vibrational wavepacket to the second order}. The calculation is performed with reduced mass 12 amu, frequency $\omega_0=1209.8\,\mathrm{cm}^{-1}$ (0.15 eV) and frequency ratio of two vibrational modes $r_1/r_2=1/3$. The modulus of density matrix elements and probability distribution for a given time $t=1.8$ fs are shown in the upper panel and lower panel, the recovered modulus and phases of density matrix elements faithfully reproduce the reference $\Pr(x_1,x_2,t)$. The algorithm converged for about 10 iterations as illustrated in the rightmost column, where $\epsilon_{10}(\hat{\rho})=4.1\times10^{-2}$ and $\epsilon_{10}(\Pr)=3.1\times10^{-2}$.}
\end{figure*}

\clearpage


\section*{Experimental data collection}

 {The details of the keV UED setup and experimental conditions for nitrogen alignment experiment have been previously introduced in \cite{Zandi:2017,Xiong:2020}. We use a tilted infrared laser pulse to excite the rotational wave packet of the nitrogen ensemble with a laser pulse duration of 60 fs,  a spot size of 190 um (horizontal) × 260 um (vertical), and pulse energy of 1mJ. The tilted angle is about 60 degrees, which is designed to remove the group velocity mismatch due to the lower speed (0.526c, where c is the speed of light) of the electron pulse. The probe electron pulse is generated by shinning a 266 nm UV laser onto a copper cathode, which is accelerated by a 90 keV DC voltage and then compressed by a 3GHz RF electric field to minimize the temporal pulse duration on the sample.  The electron beam is truncated using a platinum aperture with a diameter of 100 um to deliver a beam current of 8 pA, corresponding to 10,000 electrons per pulse. A de Laval nozzle with an inner diameter of 30 um is used to deliver the nitrogen molecules to the interaction as a supersonic molecular beam with a diameter of 200 um, and the nozzle backing pressure is 1200 mbar of nitrogen. The instrument response time was determined to be 240 fs by fitting the experimental anisotropy to its corresponding simulation. The timing jitter was 50 fs rms over several hours\cite{Xiong:2020}. The electron diffraction patterns are recorded by an electron-multiplying charge-coupled
device (EMCCD) camera, and the time delay between the pump and probe is controlled by an optical stage. Here the step of time delay is 100 fs.}
\section*{Diffraction pattern treatment}

 {

The details of how to retrieve the angular distribution from the measured diffraction patterns have been explained in \cite{Xiong:2020}. Briefly, the diffraction difference pattern for each image is calculated with $\Delta I(\textbf{s},t)=I(\textbf{s},t)-I(\textbf{s},t<-1\text{ps})$ to remove the background of atomic scattering, and then are averaged over the four quadrants using its symmetry. The simulated random molecular scattering with a rescaling factor of 0.35, which is obtained by fitting the experimental anisotropy evolution and its corresponding simulation, is added to $\Delta I(\textbf{s},t)$ to recover molecular diffraction intensity $I(\textbf{s},t)$. The modified pair distribution function (MPDF) \cite{Xiong:2020} is calculated by applying the inverse Fourier transform of $I(\textbf{s},t)$,  followed by an Abel inversion, giving the information of angular distribution $\Pr(\theta,\phi,t)$. 

The angular distribution retrieved from experimental data covers the initial alignment through the revivals up to about 7 ps,  which is deconvolved using the algorithm in \cite{Holmes:1995,Biggs:1997,Jansson:1997}. The point spread function (PSF) is assumed to be a one-dimensional Gaussian function with a full width at half maximum of 0.28 ps for the deconvolution, which eliminates the blurring due to the limit temporal resolution of the setup.  The temporal evolution of $\Pr(\theta,\phi,t)$ is extended to obtain the data up to 11ps by a reflection of the angular distribution evolution from 6.1ps to 1.2 ps to approximate the data from 6.1 ps to 11 ps according to the approximate symmetry based on the simulation. }

%
\section*{Quantum tomography for states in $m$-block with fixed projection quantum numbers}
We extend the treatment in Ref.~\cite{Mouritzen06:JCP244311} to show that the density matrix element $\denmat{J_1m_1}{J_2m_2}$ in the $(m_1, m_2)$-block subspace can be solved analytically, once the blockwise probability density $\Pr_{m_1,m_2}(\theta,t)$ of given projection quantum numbers $m_1, m_2$ is determined.
We expand the blockwise probability density with eigenbasis,
\begin{eqnarray}\label{Eq:prm}
    \Pr_{m_1,m_2}(\theta,t)=\sum_{J_1=|m_1|}^\infty \sum_{J_2=|m_2|}^\infty \denmat{J_1m_1}{J_2m_2} \widetilde{P}_{J_1}^{m_1}(\cos\theta)\widetilde{P}_{J_2}^{m_2}(\cos\theta) e^{-i\Delta\omega t}
    \,,
\end{eqnarray}
where the energy level difference is
\begin{eqnarray}
\Delta\omega=\omega_{J_1}-\omega_{J_2}=\frac{\Delta J(J+1)}{2\mathcal{I}}\nonumber
\,,
\end{eqnarray}
$\Delta J= J_1-J_2, J=J_1+J_2$ and $\mathcal{I}$ is the moment of inertia of the rotating molecule.
For the sake of convenience, we define normalized associated Legendre polynomials
\begin{eqnarray}\label{Eq:normPJm}
    \widetilde{P}_{J}^{m}(\cos\theta)=(-1)^m\sqrt{\frac{(2J+1)(J-m)!}{2(J+m)!}} {P}_{J}^{m}(\cos\theta)
    \,,
\end{eqnarray}
with orthonormal relations 
\begin{eqnarray}
    \int_0^\pi \sin\theta d\theta\widetilde{P}_{J_1}^{m}(\cos\theta) \widetilde{P}_{J_2}^m(\cos\theta)=\delta_{J_1,J_2}
    \,.
\end{eqnarray}
We use the orthogonal relations of Legendre polynomials and exponential functions in the integral transformation~\cite{Mouritzen06:JCP244311}.
Firstly, consider the motion along rotational polar coordinate $\theta$. The product of two associated Legendre polynomials occur in Eq.~\ref{Eq:prm} can be expanded by single associated Legendre polynomials
\begin{eqnarray}
    \widetilde{P}_{J_1}^{m_1}(\cos\theta)\widetilde{P}_{J_2}^{m_2}(\cos\theta)=\sum_{L=|J_1-J_2|}^{J_1+J_2} C_{J_1m_1J_2m_2}^{L,m_1+m_2} \widetilde{P}_L^{m_1+m_2}(\cos\theta)
    \,,\\
    C_{J_1m_1J_2m_2}^{L,m_1+m_2}=\sqrt{\frac{(2J_1+1)(2J_2+1)}{4\pi(2L+1)}}\IP{J_1m_1J_2m_2}{L(m_1+m_2)}\IP{J_10J_20}{L0}
    \,.
\end{eqnarray}
Thus, integrate over $\theta$,
\begin{eqnarray}
    I_{m_1m_2}(\alpha,t)&=&\int_0^\pi \sin\theta d\theta \widetilde{P}_{\alpha}^{m_1+m_2}(\cos\theta)\Pr_{m_1,m_2}(\theta,t)\\\nonumber
    &=&\sum_{J_1=|m_1|}^\infty \sum_{J_2=|m_2|}^\infty\sum_{L=|\Delta J|}^J C_{J_1m_1J_2m_2}^{L,m_1+m_2} \denmat{J_1m_1}{J_2m_2} e^{-i\Delta\omega t} \\\nonumber
    &&\times \int_0^\pi \sin\theta d\theta \widetilde{P}_{\alpha}^{m_1+m_2}(\cos\theta) \widetilde{P}_L^{m_1+m_2}(\cos\theta)\\\nonumber
    &=&\sum_{J_1=|m_1|}^\infty \sum_{J_2=|m_2|}^\infty C_{J_1m_1J_2m_2}^{\alpha,m_1+m_2} \denmat{J_1m_1}{J_2m_2} e^{-i\Delta\omega t}
    \,.
\end{eqnarray}
Let $T=4\pi \mathcal{I}$, which is related to the rotational period, and integrate over $t$,
\begin{eqnarray}
    I_{m_1m_2}(\alpha,\beta)&=& \frac{1}{T}\int_0^{T} I_{m_1m_2}(\alpha,t)e^{i\beta(\alpha+1)t/2\mathcal{I}} dt\\\nonumber
    &=&\sum_{J_1=|m_1|}^\infty \sum_{J_2=|m_2|}^\infty C_{J_1m_1J_2m_2}^{\alpha,m_1+m_2} \denmat{J_1m_1}{J_2m_2} \delta_{\beta(\alpha+1)-\Delta J(J+1)}
    \,.
\end{eqnarray}
The range of $\alpha$ and $\beta$ is set to be $|\Delta J|\le |\beta| \le\alpha\le J$, where $\beta$ and $\Delta J$ are of the same sign. 
If $\beta(\alpha+1)$ has unique integer factorization, the only term remaining in the sum satisfying
\begin{eqnarray}
\beta(\alpha+1)=\Delta J(J+1)
\end{eqnarray}
is $\beta=\Delta J$ and $\alpha=J$. The corresponding density matrix element can be derived as 
\begin{eqnarray}
\denmat{\frac{\alpha+\beta}{2}m_1}{\frac{\alpha-\beta}{2}m_2}=\frac{I_{m_1m_2}(\alpha,\beta)}{C_{\frac{\alpha+\beta}{2}m_1\frac{\alpha-\beta}{2}m_2}^{\alpha,m_1+m_2} }
\,.
\end{eqnarray}
If the factorization of $\beta(\alpha+1)$ is not unique, we calculate all integrations $I_{m_1m_2}(\alpha',\beta')$ where $\beta(\alpha+1)=\beta'(\alpha'+1)$.  For example, when $\beta=0$, 
\begin{eqnarray}
    I_{m_1m_2}(\alpha,0)=\sum_{J=\mathrm{max}\{|m_1|,|m_2|\}}^\infty  C_{Jm_1Jm_2}^{\alpha,m_1+m_2} \denmat{Jm_1}{Jm_2}
\end{eqnarray}
all of the $\Delta J=0$ terms remain. When changing the value of $\alpha$, all these $I_{m_1m_2}$ and corresponding density matrix elements constitute a set of linear algebraic equations (where $\alpha=2J$ can only be even numbers),
\begin{eqnarray}
\left(\begin{array}{c}
     I_{m_1m_2}(\alpha,0)  \\
     I_{m_1m_2}(\alpha+2,0) \\
     I_{m_1m_2}(\alpha+4,0) \\
     \vdots
\end{array}\right)
=\left(\begin{array}{cccc}
     C_{\frac{\alpha}{2}m_1\frac{\alpha}{2}m_2}^{\alpha,m_1+m_2} & C_{\frac{\alpha}{2}+1,m_1,\frac{\alpha}{2}+1,m_2}^{\alpha,m_1+m_2} & C_{\frac{\alpha}{2}+2,m_1,\frac{\alpha}{2}+2,m_2}^{\alpha,m_1+m_2} & \cdots\\
    0 & C_{\frac{\alpha}{2}+1,m_1,\frac{\alpha}{2}+1,m_2}^{\alpha+2,m_1+m_2} & C_{\frac{\alpha}{2}+2,m_1,\frac{\alpha}{2}+2,m_2}^{\alpha+2,m_1+m_2}& \cdots\\
    0 & 0 & C_{\frac{\alpha}{2}+2,m_1,\frac{\alpha}{2}+2,m_2}^{\alpha+4,m_1+m_2}& \cdots\\
    \vdots& \vdots& \vdots& \cdots
\end{array}\right)\\\nonumber
\times
\left(\begin{array}{c}
    \denmat{\frac{\alpha}{2}m_1}{\frac{\alpha}{2}m_2} \\
    \denmat{\frac{\alpha}{2}+1,m_1}{\frac{\alpha}{2}+1,m_2} \\
    \denmat{\frac{\alpha}{2}+2,m_1}{\frac{\alpha}{2}+2,m_2} \\
    \vdots
\end{array}\right)
\,,
\end{eqnarray}
which has unique solution because all diagonal terms of the upper triangular matrix are nonzero.

\section*{Laser alignment of rotating molecule}
The effective Hamiltonian of rotating molecule-laser interaction is~\cite{Stapelfeldt03:RMP543}
\begin{eqnarray}\label{Eq:lasHam}
&&\hat{H}_{\mathrm{eff}}=\hat{H}_0+\hat{H}_{\mathrm{int}}\nonumber\\
&&\hat{H}_0=B\textbf{J}^2\nonumber\\
&&\hat{H}_{\mathrm{int}} = -\frac12\epsilon^2(t)[(\alpha_{\parallel}-\alpha_{\perp})\cos^2\theta+\alpha_{\perp}]
\,,
\end{eqnarray}
where $\textbf{J}$ is the rotational angular momentum, $\epsilon(t)$ is the electric field of the laser pulse, $B$ is the rotational constant, $\alpha_{\parallel}$ and $\alpha_{\perp}$ are the components of the static polarizability, parallel and perpendicular to the molecular axes.
The molecule is assumed to be in the vibrational and electronic ground state. 
An initial rotational eigenstate $\ket{J_0M_0}$ evolves to a pendular state~\cite{Stapelfeldt03:RMP543}
\begin{eqnarray}\label{Eq:Norm_ofd}
\ket{J_0m_0}\rightarrow \ket{\psi(t)}^{(J_0m_0)}=
\sum_{J}d_J^{(J_0m_0)}\ket{Jm_0}e^{-iE_{J}t/\hbar}
\,,
\end{eqnarray}
where $J$ and $J_0$ are of the same parity. The coupling coefficients $d_J^{J_0m_0}$ is induced by laser field, satisfying selection rules $\Delta m=0$ and $\Delta J=0,\pm 2$. $d_J^{J_0m_0}$ is invariant after the laser pulse, and the evolution of rotational angular distribution originates from interference of each dynamical phase.
The coherence of the created quantum state can be maintained for several revival periods, and the alignment is reconstructed at predetermined times and survives for a perfectly controllable period~\cite{Stapelfeldt03:RMP543}, the sufficiently long coherence time makes the time evolution measurement of quantum state tomography feasible.

The initial system in thermal equilibrium can be characterized by the following density operator
\begin{eqnarray}
  \hat{\rho}_{\mathrm{ini}} &=& \sum_{J_0m_0}\omega_{J_0} \ketBra{J_0m_0}{J_0m_0}
  \,,
\end{eqnarray}
where $\omega_{J_0}$ is the Boltzmann statistical factor determined by the rotational temperature. 
The density operator of the laser-aligned system is
\begin{eqnarray}
 \hat{\rho}(t) &=& \sum_{J_0m_0}\omega_{J_0} \ketBra{\psi(t)^{(J_0m_0)}}{\psi(t)^{(J_0m_0)}}\\\nonumber
 &=&\sum_{m_0} \left[\sum_{J_0}\omega_{J_0} \left(\sum_{J_1}d_{J_1}^{(J_0m_0)}\ket{J_1m_0}\right)\left(\sum_{J_2}d_{J_2}^{*(J_0m_0)}\Bra{J_2m_0}\right)\right]e^{-i(E_{J_1}-E_{J_2})t/\hbar}\\\nonumber
 &=&\sum_{J_1J_2m}\left(\sum_{J_0}\omega_{J_0} d_{J_1}^{(J_0m)}d_{J_2}^{*(J_0m)}\right)e^{-i(E_{J_1}-E_{J_2})t/\hbar}\ketBra{J_1m}{J_2m}
 \,.
\end{eqnarray}
And its density matrix elements are
\begin{eqnarray}
\BK{J_1m_1}{\hat{\rho}(t)}{J_2m_2}=\delta_{m_1m_2}\left(\sum_{J_0}\omega_{J_0} d_{J_1}^{(J_0m_1)}d_{J_2}^{*(J_0m_2)}\right)e^{-i(E_{J_1}-E_{J_2})t/\hbar}
\,.
\end{eqnarray}
So the partial trace of $m$ subspace with odd (or even) $J$ is invariant in the dynamics of laser alignment, since it is a general property of laser-molecule interaction,
\begin{eqnarray}\label{Eq:prconst}
\sum_{J_{\odd}}\BK{Jm}{\hat{\rho}}{Jm}=
 \sum_{J_{\odd}}\sum_{J_{0\,\odd}}\omega_{J_0} |d_J^{(J_0m)}(t)|^2=\sum_{J_{0\,\odd}}\omega_{J_0}
 \,,
\end{eqnarray}
where we used the normalization property of coefficients $d_J^{J_0M}(t)$ in Eq.~\ref{Eq:Norm_ofd}.

Notice that density matrix of opposite magnetic quantum number $m$ and $-m$ is symmetric for $\hat{\rho}_{\mathrm{ini}}$, which also remains symmetric for transition matrix element induced by laser interaction $\hat{H}_{\mathrm{eff}}(t)$. From Eq.~\ref{Eq:lasHam}, taking into account selection rule $\Delta M=0$,
\begin{eqnarray}\nonumber
  &&\BK{J_1m}{\hat{H}_{\mathrm{eff}}(t)}{J_2m}=\BK{J_1,-m}{\hat{H}_{\mathrm{eff}}(t)}{J_2,-m}\\
  &=&\delta_{J_1,J_2}\left[BJ_1(J_1+1)-\frac12\epsilon^2(t)\alpha_{\perp}\right]-\frac12\epsilon^2(t)(\alpha_{\parallel}-\alpha_{\perp})\BK{J_1m}{\cos^2\theta}{J_2m}
  \,,
\end{eqnarray}
where $\BK{J_1m}{\cos^2\theta}{J_2m}=\BK{J_1,-m}{\cos^2\theta}{J_2,-m}$ according to the properties of Clebesh-Gordan coefficients. The coefficients of pendular state $d_{J}^{(J_0m_0)}$, which are totally determined by initial condition $\hat{\rho}_{\mathrm{ini}}$ and the Schr\"{o}dinger equation,
\begin{eqnarray}
   i\dot{d}_{J}^{(J_0m)}=\sum_{J'}\BK{Jm}{\hat{H}_{\mathrm{eff}}(t)}{J'm}
   \,,
\end{eqnarray}
are also symmetric ${d}_{J}^{(J_0m)}={d}_{J}^{(J_0,-m)}$. So are the density matrix elements
\begin{eqnarray}
   \denmat{J_1m_1}{J_2m_2}=\sum_{J_0}\omega_{J_0}{d}_{J_1}^{(J_0,m_1)}{d}_{J_2}^{*(J_0,m_2)}=\denmat{J_1, -m_1}{J_2, -m_2}
   \,.
\end{eqnarray}
\section*{The algorithm for imposing constraints of iterative quantum tomography}
In this section we show the detailed procedure for making an arbitrary density matrix and probability distribution to satisfy the physical constraints given in the main text. 
Most physical constraints are given in the summation form. For example, from Eq.~\ref{Eq:prconst},
\begin{eqnarray}
\sum_{J_{\odd}}\BK{Jm}{\hat{\rho}}{Jm}=\sum_{J_{0\,\odd}}\omega_{J_0}
\,.
\end{eqnarray}
From the measured probability distribution
\begin{eqnarray}
&&\widetilde{\Pr}_{m_1-m_2}(\theta,t)=\int_0^{2\pi}d\phi \Pr(\theta,\phi,t) e^{-i(m_1-m_2)\phi} \\\nonumber
&=& \frac{1}{2\pi}\sum_{J_1m_1'J_2m_2'} \denmat{J_1m_1'}{J_2m_2'} \widetilde{P}_{J_1}^{m_1}(\cos\theta)\widetilde{P}_{J_2}^{m_2}(\cos\theta)e^{-i\Delta\omega t} \int_0^{2\pi}d\phi e^{im_1'\phi}e^{-im_2'\phi}e^{-i(m_1-m_2)\phi}\\\nonumber
&=&\sum_{m_1'm_2'}\delta_{m_1'-m_2',m_1-m_2}\sum_{J_1J_2}\denmat{J_1m_1'}{J_2m_2'} \widetilde{P}_{J_1}^{m_1}(\cos\theta)\widetilde{P}_{J_2}^{m_2}(\cos\theta)e^{-i\Delta\omega t}
\,,
\end{eqnarray}
and the constraint can be expressed as
\begin{eqnarray}
\sum_{m_1'-m_2'=m_1-m_2} \Pr_{m_1',m_2'}(\theta,t)=\widetilde{\Pr}_{m_1-m_2}(\theta,t)
\,.
\end{eqnarray}
They can be sataisfied by scaling with a common factor
\begin{eqnarray}\label{Eq:alp}
\BK{Jm}{\hat{\rho}}{Jm}\rightarrow \alpha\BK{Jm}{\hat{\rho}}{Jm}\,,&\quad& \alpha=\frac{\sum_{J_{0\,\odd}}\omega_{J_0}}{\sum_{J_{\odd}}\BK{Jm}{\hat{\rho}}{Jm}}\,.\\\label{Eq:Pr}
\Pr_{m_1,m_2}(\theta,t)\rightarrow \beta(\theta,t)\Pr_{m_1,m_2}(\theta,t)\,, &\quad& \beta=\frac{\widetilde{\Pr}_{m_1-m_2}(\theta,t)}{\sum_{m_1'-m_2'=m_1-m_2} \Pr_{m_1',m_2'}(\theta,t)}
\,.
\end{eqnarray}
The constraints in probability space is given by Eq.~\ref{Eq:Pr}, and illustrated with flow chart in Fig.~\ref{fig:constraint_Pr}. Further constraints in density matrix space include being Hermitian, positive semidefinite and having invariant partial traces (the procedure is presented with the flow chart in Fig.~\ref{fig:constraint_dm}).

As a general rule to guarantee the completeness of constraint conditions, we can firstly analyse the physical system and find out the possible states, which could give same probability distribution for all time and are indistinguishable without further constraint, and construct the set of physical conditions that can distinguish the states from each other, e.g. selection rules, symmetry. The obtained physical conditions can be then used as constraints in the iterative QT procedure. In this manner, the completeness of the constraint conditions and the faithfulness of the converged density matrix solution can be achieved, i.e. the converged solution of the inversion problem is the true density matrix of the physical system.

\section*{Benchmarking iterative quantum tomography with simulated ultrafast diffraction of N$_2$ rotational wavepacket}

We use the new QT method to extract rotational density matrix from simulated ultrafast diffraction dataset of impulsively aligned nitrogen molecule, prepared at rotational temperature of 30 K.
As shown in Fig.~\ref{fig:diffpic_sim}, from a simulated dataset consisting of a series of time-ordered snapshots of diffraction patterns~\cite{HO08:PRA052409}
\begin{eqnarray}
I(\textbf{s},t)=\int_0^{2\pi}d\phi\int_{0}^{\pi}\sin\theta d\theta \Pr(\theta,\phi,t) |f(\textbf{s},\theta,\phi)|^2
\,,
\end{eqnarray}
the time-dependent molecular probability distribution $\mathrm{Pr}(\theta,\phi,t)$ can be obtained by solving the Fredholm integral equation of the first kind using Tikhonov regularization procedure~\cite{Ischenko14:Adv}. We assume $\tau=-\cos\theta$ and replace the integral by Riemann summation,
\begin{eqnarray}\label{Eq:Fredh}
I(\Theta_{k},\Phi_{l})=\sum_{i=1}^{a}\Delta\phi\sum_{j=1}^{b}\Delta \tau|f(\phi_{i},\theta(\tau_{j}),\Theta_k,\Phi_l)|^2\Pr(\phi_{i},\theta(\tau_{j}))\,,
\end{eqnarray}
at each instant, where $\Delta\phi=\frac{2\pi}{a}$, $\Delta \tau =\frac{2}{b}$, $i$ is ranging from $1$ to $a$, $j$ is ranging from $1$ to $b$, $k$ is ranging from $1$ to $c$, and $l$ is ranging from $1$ to $d$. $\phi$ and $\theta$ are the azimuthal and levitation angles of the linear molecular rotor, $\Theta$ and $\Phi$ are the scattering angle of the X-ray photon in the lab system (as is shown in Fig.~1 in the main text).
We can write the total diffraction intensity in the matrix form $I=\mathbf{K}\Pr$,
where 
\begin{eqnarray}
&&I={
\left( \begin{array}{c}
I(\Theta_{1},\Phi_{1})\\
\vdots\\
I(\Theta_{1},\Phi_{d})\\
I(\Theta_{2},\Phi_{1})\\
\vdots\\
I(\Theta_{c},\Phi_{d})
\end{array} 
\right )},\nonumber\\
&&\mathbf{K}={
\left( \begin{array}{ccc}
|f(\phi_{1},\theta_{1},\Theta_{1},\Phi_{1})|^{2}\Delta\phi\Delta \tau& \cdots&
|f(\phi_{a},\theta_{b},\Theta_{1},\Phi_{1})|^{2}\Delta\phi\Delta \tau\\
\vdots& \ddots & \vdots\\
|f(\phi_{1},\theta_{1},\Theta_{c},\Phi_{d})|^{2}\Delta\phi\Delta \tau & \cdots & |f(\phi_{a},\theta_{b},\Theta_{c},\Phi_{d})|^{2}\Delta\phi\Delta \tau
\end{array}
\right )},\nonumber\\
&&\Pr={
\left( \begin{array}{c}
\Pr(\phi_{1},\theta_{1})\\
\vdots\\
\Pr(\phi_{1},\theta_{b})\\
\Pr(\phi_{2},\theta_{1})\\
\vdots\\
\Pr(\phi_{a},\theta_{b})
\end{array} 
\right )}
\,.
\end{eqnarray}
To avoid singular matrix inversion, we use Tikhonov regularization to get the rotational probability distribution,
\begin{eqnarray}
\Pr=(\mathbf{K}^{T}\mathbf{K}+\lambda \mathbf{E})^{-1}\mathbf{K}^{T}I 
\,,
\end{eqnarray}
where $\mathbf{E}$ is identity matrix of size $(c\times d)$ and $\mathbf{K}^{T}$ is the transpose of matrix $\mathbf{K}$.

The Tikhonov regularization performs excellently in dealing with experimental data with measurement errors and preventing overfitting, and can faithfully recover the probability density distribution.
To validate the faithfulness of the obtained probability distribution $\Pr(\theta,\phi)$, we define the condition number
\begin{eqnarray}
\mathrm{cond}=\frac{\|\Delta{\mathrm{Pr}}\|_{2}/\|\mathrm{Pr}\|_{2}}{\|\Delta{I}\|_{2}/\|I\|_{2}}
\,,
\end{eqnarray}
where $\|A\|_{2}=\sqrt{\sum_{i}A_{i}^{2}}$ is the $\mathcal{L}^2$ Euclid norm.
The condition number characterizes the degree of variation of the solution $\Pr(\theta,\phi)$ with respect to the input data of measured diffraction intensity $I(\textbf{s})$, its value provides a measure for the sensitivity of the solution with respect to the measurement error and choice of regularization parameters. From Fig.~\ref{fig:Sensitivity_Analysis}, we can estimate that $\lambda\geq 10$ is required to ensure $\mathrm{cond}\leq 10$, and subsequently to ensure the reliability of the solution. 

Quantum tomography of the rotational wavepacket gives the result shown in Fig.~3 in the main text. After 50 iterations, both density matrix and probability distribution are precisely recovered. 
The error of density matrix is $\epsilon_{50}(\hat{\rho})=2.9\times 10^{-2}$ and error of probability achieves $\epsilon_{50}(\Pr)=3.8\times 10^{-5}$.
\section*{Numerical trial with randomly chosen density matrix and initial guess}
We have verified the new quantum tomographic method by the rotational wavepacket of a laser-aligned molecule. We also illustrate the power of the new method by applying it to {a randomly chosen} density matrix rather than that in the laser-aligned case. The iterative QT algorithm also converges after about 20 iterations and density matrix is recovered with considerable accuracy.
The density operator of the state to be recovered is set to be
\begin{eqnarray}\nonumber   
\hat{\rho}&=&\frac{2}{21}\ketBra{00}{00}+\frac{3}{14}\ketBra{10}{10}+\frac{1}{42}\ketBra{20}{20}\nonumber\\
 &+& \left(\frac{1}{7}\ketBra{00}{10}+\frac{1}{21}\ketBra{00}{20}+\frac{1}{14}\ketBra{10}{20}+\mathrm{H.c.}\right)\nonumber\\
 &+&\frac{1}{21}\ketBra{11}{11}+\frac{3}{28}\ketBra{21}{21}+\frac{1}{84}\ketBra{31}{31}\nonumber\\
 &+& \left(\frac{1}{14}\ketBra{11}{21}+\frac{1}{42}\ketBra{11}{31}+\frac{1}{28}\ketBra{21}{31}+\mathrm{H.c.}\right)\nonumber\\
 &+&\frac{1}{21}\ketBra{1,-1}{1,-1}+\frac{3}{28}\ketBra{2,-1}{2,-1}+\frac{1}{84}\ketBra{3,-1}{3,-1}\nonumber\\
 &+& \left(\frac{1}{14}\ketBra{1,-1}{2,-1}+\frac{1}{42}\ketBra{1,-1}{3,-1}+\frac{1}{28}\ketBra{2,-1}{3,-1}+\mathrm{H.c.}\right)\nonumber\\
 &+&\frac{1}{21}\ketBra{22}{22}+\frac{3}{28}\ketBra{32}{32}+\frac{1}{84}\ketBra{42}{42}\nonumber\\
 &+& \left(\frac{1}{14}\ketBra{22}{32}+\frac{1}{42}\ketBra{22}{42}+\frac{1}{28}\ketBra{32}{42}+\mathrm{H.c.}\right)\nonumber\\
 &+&\frac{1}{21}\ketBra{2,-2}{2,-2}+\frac{3}{28}\ketBra{3,-2}{3,-2}+\frac{1}{84}\ketBra{4,-2}{4,-2}\nonumber\\
 &+& \left(\frac{1}{14}\ketBra{2,-2}{3,-2}+\frac{1}{42}\ketBra{2,-2}{4,-2}+\frac{1}{28}\ketBra{3,-2}{4,-2}+\mathrm{H.c.}\right)
 \,.
\end{eqnarray}
We impose the error functions of density matrix and probability distribution to measure the accuracy of iteration results, which are defined by
\begin{eqnarray}
\epsilon_{n}(\hat{\rho})&=&\frac{\sum_{J_1m_1J_2m_2}|\denmat{J_1m_1}{J_2m_2}_n-\denmat{J_1m_1}{J_2m_2}_0|}{\sum_{J_1m_1J_2m_2}|\denmat{J_1m_1}{J_2m_2}_0|}\\
\epsilon_{n}(\Pr)&=&\frac{\sum_{i,j,k}|\Pr_n(\theta_i,\phi_j,t_k)-\Pr_0(\theta_i,\phi_j,t_k)|}{\sum_{i,j,k}|\Pr_0(\theta_i,\phi_j,t_k)|}   
\end{eqnarray}
where the subscript $n$ represents the result of $n$-th iteration, and $0$ represents the correct result. 

In Fig.~\ref{Fig:NumTrial} we show the result of identical algorithm given in Fig.~\ref{fig:constraint_Pr} and Fig.~\ref{fig:constraint_dm}, only with smaller order $J_{\mathrm{max}}$ of density matrix to be recovered. 
 {The initial state is given by correct diagonal elements of density matrix.} The iteration converged to the expected result with error $\epsilon_{20}(\hat{\rho})=3.5\times10^{-3}$ and $\epsilon_{20}(\Pr)=1.7\times10^{-3}$. 

Especially, we show with the proof-of-principle example that this iterative QT algorithm is insensitive with the initial guess of density matrix.
The rotational temperature which provides much information such as initial guess and partial trace, is actually not indispensable to the QT method. Assume we are dealing with a pure QT problem without any additional knowledge to the density matrix to be recovered. As is shown in Fig.~\ref{Fig:rdm}, a random initial guess will also lead to a converged result after about 30 iterations with error $\epsilon_{30}(\hat{\rho})=3.9\times10^{-2}$ and $\epsilon_{30}(\Pr)=9.0\times10^{-3}$.

\section*{Vibrational and electronic quantum tomography}
Vibrational quantum tomography recovers the density matrix of $N$ vibrational modes from the probability distribution evolution $\Pr(x_1,x_2,\cdots,x_N,t)$
\begin{eqnarray}\label{denmat2prb}
    \Pr(x_1,x_2,\cdots,x_N,t)&=&\sum_{\{m_i\}_{i=1}^{N}}\sum_{\{n_i\}_{i=1}^{N}}\denmat{n_1n_2\cdots n_N}{m_1m_2\cdots m_N} \\\nonumber
    &\times& \prod_{i=1}^{N}\phi_{n_i}(x_i)\phi^*_{m_i}(x_i)e^{i(m_i-n_i)\omega_i t}
    \,.
\end{eqnarray}
where $\phi_{n_i}(x_i)$ is the harmonic oscillator wavefunction of the $i$-th vibrational mode with energy eigenvalue $(n_i+\frac{1}{2})\omega_i$. The dimension problem arises naturally. Here the probability is $(N+1)$-dimensional and density matrix is $2N$-dimensional, which is inadmissible for analytical solutions when $N>1$. In conventional QT method that is based on integral transform, the orthogonal properties cancel out one summation by integrating over one parameter. For example,
\begin{eqnarray}
\frac{1}{T}\int_{0}^{T}dt e^{i(m-n)r\omega_0 t} e^{-ik\omega_0 t}=\delta_{(m-n)r,k}
\,,
\end{eqnarray}
where $T=\frac{2\pi}{\omega_0}$. $f_{mn}(x)$ is the sampling function~\cite{Leonhardt96:PRL1985} defined by
\begin{eqnarray}
f_{mn}(x)=\frac{\partial}{\partial x}[\phi_{m}(x)\varphi_{n}(x)]
\,,
\end{eqnarray}
where $\phi_m(x)$ and $\varphi_n(x)$ are respectively regular and irregular wavefunctions of harmonic oscillator. The bi-orthogonal properties of sampling function is
\begin{eqnarray}
\int_{-\infty}^{+\infty}dx f_{mn}(x) \phi^*_{m'}(x) \phi_{n'}(x)&=&\delta_{mm'}\delta_{nn'}
\,,
\end{eqnarray}
under frequency constraints $m-n=m'-n'$.

Our theory, based on the following two procedures, fully utilizes the above orthogonal properties and imposes constraints for lack of dimension. First, we set up the transformation between probability and density matrix in a subspace
\begin{eqnarray}
    \Pr_{\Delta_1,\Delta_2,\cdots,\Delta_N}(x_1,x_2,\cdots,x_N)&=&\sum_{\{m_i\}_{i=1}^{N}}\sum_{\{n_i\}_{i=1}^{N}}\denmat{n_1n_2\cdots n_N}{m_1m_2\cdots m_N}\\\nonumber
    &&\times\prod_{i=1}^{N}\phi_{n_i}(x_i)\phi^*_{m_i}(x_i)\delta_{m_i-n_i,\Delta_i}\\
    \denmat{n_1n_2\cdots n_N}{m_1m_2\cdots m_N}&=&
    \int d^N\textbf{x} \Pr_{\Delta_1,\Delta_2,\cdots,\Delta_N}(x_1,x_2,\cdots,x_N)
    \prod_{i=1}^{N}f_{m_in_i}(x_i)\label{Eq:convolv_pattern_func}
    \,.
\end{eqnarray}
Second, starting from an initial guess, effective physical constraints can be imposed by iterative projection method to get the converged result. For example, the priori knowledge of density matrix of being Hermitian, positive semidefinite and normalized. The algorithm of vibrational state QT and an example of 2D vibrational quantum tomography is shown in Fig.~\ref{fig:vib} and Fig.~\ref{fig:vibqtres}. The initial guess is given randomly, and only the probability distribution and general properties of density matrix are imposed as constraints during the iteration algorithm.

Similar to rotational QT, the dimension problem can be reflected by the fact that for
\begin{eqnarray}\label{dimp}
\Pr_k(x_1,x_2,\cdots,x_N)=\sum_{\{\Delta_i\}_{i=1}^{N}} \Pr_{\Delta_1,\Delta_2,\cdots,\Delta_N}(x_1,x_2,\cdots,x_N)\delta_{\sum_{i=1}^{N}\Delta_ir_i,k}
\,,
\end{eqnarray}
unless only one single combination of $\{\Delta_i\}$ satisfies $\sum_{i=1}^{N}\Delta_ir_i=k$, there is no direct way to obtain $\Pr_{\Delta_1,\Delta_2,\cdots,\Delta_N}(x_1,x_2,\cdots,x_N)$ from the measured wavepacket density distribution, only their sum can be available through Fourier transform of the measured probability distribution evolution
\begin{equation}\label{Eq:tmres}
    \Pr_k(x_1,x_2,\cdots,x_N)=\frac{1}{T}\int_{0}^{T}dt e^{-ik\omega_0 t} \Pr(x_1,x_2,\cdots,x_N,t)
    \,,
\end{equation}
where we assume $\omega_i=r_i\omega_0$ ($r_i$ are integers and $T=2\pi/\omega_0$, $r_i$'s are the set of smallest integers to represent the measured frequencies).
In the new iterative QT method for $N$-dimensional vibrational system, we do not need infinitely long time of measurement anymore, which used to be indispensable to fill the whole space of $N$-dimensional phases~\cite{Mouritzen05:PRA} while physically infeasible. 
Besides, in the new iterative QT method, the ratio of frequencies does not have to be irrational, which is important because in reality $N$-dimensional vibrational systems with commensurable frequencies are ubiquitous.

The pattern function can be approximated around $x=0$ as~\cite{Leonhardt96:Opt}
\begin{eqnarray}
f_{nn}\sim -\frac{2}{\pi}\sin[-\pi(n+1/2)+2\sqrt{2n+1}x]
\,.
\end{eqnarray}
In order to resolve a period of the oscillation of the pattern function that arises in the convolution (Eq.~\ref{Eq:convolv_pattern_func}), the required spatial resolution for reconstructing vibrational density matrix up to $N$-th order has to be better than $\delta{x}\leq \pi/2\sqrt{2N+1}$. 
The maximal order of the desired density matrix also sets demand on the temporal resolution. Suppose $d$ time intervals are measured for a half period ${T}/{2}={\pi}/{\omega_0}$. 
From Eq.~\ref{Eq:tmres}, we have a phase resolution of $k\pi/d$ for the Fourier transformation of probability distribution function. The aliasing phenomena defines the maximal order of density matrix we can access to be $N=d/k-1$, thus the required temporal resolution is 
\begin{eqnarray}
\delta{t}\leq \frac{T}{2(N+1)k}\leq \frac{T}{2(N+1)\sum_{i}r_i}
\,.
\end{eqnarray}

The quantum tomography procedure presented above can be easily generalized to systems when coupling among different vibrational modes exist. In general case, the Hamiltonian \cite{Bowman79:JCP912}
\begin{eqnarray}
\hat{H}=\sum_{i=1}^{N}\hat{h}_i(x_i)+V(x_1,x_2,\cdots,x_N)
\,,
\end{eqnarray}
where $\hat{h}_i$ is the separable part for $i$-th vibrational mode with eigenstate $\phi_{n_i}(x_i)$, and $V(x_1,x_2,\cdots,x_N)$ is
coupling potential among $N$ vibrational modes. The eigenstate is a linear combination of product 1D wavefunctions assigned with quantum numbers $I=\{I_1,I_2,\cdots,I_N\}$ with energy eigenvalue $E_I$
\begin{eqnarray}
    \Psi_{I}(x_1,x_2,\cdots,x_N)=\sum_{i_1,i_2,\cdots,i_N}C_{I}^{i_1,i_2,\cdots,i_N}\prod_{\alpha=1}^{N}\phi_{i_{\alpha}}(x_\alpha)
    \,.
\end{eqnarray}
The iterative projection algorithm for quantum tomography should be set up based on the transformation between probability and density matrix in a subspace
\begin{eqnarray}
    \Pr_{\Delta_1,\Delta_2,\cdots,\Delta_N}(x_1,x_2,\cdots,x_N)=\sum_{I,J}\denmat{I}{J}\sum_{i_1,i_2,\cdots,i_N}\sum_{j_1,j_2,\cdots,j_N}C^{i_1,i_2,\cdots,i_N}_{I}C^{j_1,j_2,\cdots,j_N*}_{J}\\\nonumber
    \times
    \prod_{\alpha=1}^{N}\phi_{i_\alpha}(x_\alpha)\phi^*_{j_\alpha}(x_\alpha)\delta_{i_\alpha-j_\alpha,\Delta_\alpha}\\\label{Eq:conp}
    \int d^N\textbf{x} \Pr_{\Delta_1,\Delta_2,\cdots,\Delta_N}(x_1,x_2,\cdots,x_N) \prod_{\alpha=1}^{N}f_{i_\alpha j_\alpha}(x_\alpha)=\sum_{I,J}\denmat{I}{J}C^{i_1,i_2,\cdots,i_N}_{I}C^{j_1,j_2,\cdots,j_N*}_{J}
    \,.
\end{eqnarray}
where the frequency constraint of sampling function requires $i_\alpha-j_\alpha=\Delta_\alpha\, (\alpha=1,2,\cdots,N)$. The density matrix element can be solved from the linear equation of~\ref{Eq:conp}. If there are $n$ basis eigenstate for $i$-th uncoupled vibrational mode $\phi_{n_i}(x_i)$, the coupled density matrix can be recovered to the order of $(2n)^{N/2}$. Similarly, the procedure starts from an initial guess and imposes constraints to both density matrix space and probability space. Besides basic properties of density matrix and probability distribution, the subspace probability should also satisfy
\begin{eqnarray}
&&\Pr_{\omega_{IJ}}(x_1,x_2,\cdots,x_N)=\frac{1}{T}\int_0^{T} dt\Pr(x_1,x_2,\cdots,x_N,t)e^{-i\omega_{IJ}t}\\\nonumber
&=&\sum_{\omega_I-\omega_J=\omega_{IJ}}\denmat{I}{J}\varphi_{i_1,i_2,\cdots,i_N}(x_1,x_2,\cdots,x_N)\varphi^*_{j_1,j_2,\cdots,j_N}(x_1,x_2,\cdots,x_N)\\\nonumber
&=&\sum_{\Delta_1,\Delta_2,\cdots,\Delta_N}\Pr_{\Delta_1,\Delta_2,\cdots,\Delta_N}(x_1,x_2,\cdots,x_N)\delta_{\omega_I-\omega_J,\omega_{IJ}}
\,.
\end{eqnarray}
where $\omega_I$ and $\omega_J$ are energy eigenvalues of the coupled Hamiltonian, $T$ is the common period for all vibrational frequency intervals.

{To enhance the convergence of iterative QT procedure for vibrational states, physical constraints can be imposed on the diagonal matrix elements of the density matrix, which is experimentally accessible, e.g. through photoelectron spectra and absorption spectra, which can directly provide constraints on diagonal density matrix elements of basis states with eigenenergy $E$~\cite{Heller78:2066}.
} 

As a final remark, for vibrational QT, it is sometimes neccessary to use 
the velocities of nuclei as constraining physical conditions, in the case that the basis states of density matrix is energetically degenerate. For example, given the ratio of two vibrational frequencies $r_1/r_2=1/2$, consider a mixed state consisting of $\ket{20}$ and $\ket{10}$ (the pure state is a special case of it), their density matrix is
\begin{eqnarray}
{\bf \rho}=
\left(
\begin{array}{cc}
    \denmat{20}{20} & \denmat{20}{01} \\
    \denmat{01}{20} & \denmat{01}{01}
\end{array}
\right)
=
\left(
\begin{array}{cc}
    \rho_{11} & \rho_{12} \\
    \rho_{21} & \rho_{22}
\end{array}
\right)
\,.
\end{eqnarray}
The probability distribution
\begin{eqnarray}
\Pr(x_1,x_2,t)&=&\rho_{11} \phi^2_{2}(x_1)\phi^2_{0}(x_2) +\rho_{22}\phi^2_{0}(x_1)\phi^2_{1}(x_2) \\\nonumber
&+&(\rho_{12}+\rho_{21})\phi_{2}(x_1)\phi_{0}(x_2)\phi_{0}(x_1)\phi_{1}(x_2)
\,
\end{eqnarray}
could not reflect the imaginary part of the off-diagonal density matrix elements because the degeneracy of the two basis states smears out the temporal evolution of the probability distribution.
If $\ket{20}$ and $\ket{01}$ belong to the same symmetry representation, their coupling will lead to Fermi resonance and the degeneracy can be lifted.
In the case that $\ket{20}$ and $\ket{01}$ are exactly degenerate, additional constraints must be imposed.
Because with the ultrafast diffraction method, the velocity of nuclei and  thus their momenta can be extracted experimentally, we can naturally construct physical constraints through products of momenta, such as ${p}^2_{x_1}{p}_{x_2}$, since
\begin{eqnarray}
{\bf A}=(\hat{p}^2_{x_1}\hat{p}_{x_2})=\left(
\begin{array}{cc}
    a_{11} & a_{12} \\
    a_{21} & a_{22}
\end{array}
\right)
\end{eqnarray} 
has nonzero imaginary part of non-diagonal matrix elements. For example,
\begin{eqnarray}\nonumber
a_{12}&=&\int dx_1 \phi_{2}(x_1) \left(- \frac{\partial^2}{\partial x_1^2}\right) \phi_0(x_1) \int dx_2 \phi_0(x_2)  \left(-i  \frac{\partial}{\partial x_2}\right) \phi_1(x_2)\\\nonumber
&=& \int_{-\infty}^{\infty} dx_1 \frac{1}{\pi^{1/4}}\sqrt{\frac{\alpha_1}{2}}(2\alpha_1^2x_1^2-1)e^{-\frac{1}{2}\alpha_1x_1^2}\frac{\partial^2}{\partial x_1^2}\left(\frac{\sqrt{\alpha_1}}{\pi^{1/4}}e^{-\frac{1}{2}\alpha_1x_1^2}\right)\\\nonumber
&&\times \int_{-\infty}^{\infty} dx_2 \frac{\sqrt{\alpha_2}}{\pi^{1/4}}e^{-\frac{1}{2}\alpha_2x_2^2} \frac{\partial}{\partial x_2}\left(\frac{\sqrt{2\alpha_2}}{\pi^{1/4}}e^{-\frac{1}{2}\alpha_2x_2^2}\right) = -i \frac{\alpha_1^2\alpha_2}{2}  \\
a_{21}&=&a^*_{12}=i \frac{\alpha_1^2\alpha_2}{2}
\end{eqnarray}
The observable
\begin{eqnarray}
\langle\hat{A}\rangle&=& m_1^2v^2_{1}m_2v_{2}=\Tr (\hat{\rho}\hat{A}) \\\nonumber
&=& \rho_{11} a_{11}+\rho_{12}a_{21} +\rho_{21}a_{12}+\rho_{22}a_{22} \\\nonumber
&=& \rho_{11} a_{11}+\rho_{22}a_{22}+2\Re[\rho_{12}a_{21}]\\\nonumber
&=&\rho_{11} a_{11}+\rho_{22}a_{22}-\alpha_1^2\alpha_2 \Im[\rho_{12}]
\end{eqnarray}
contains information of imaginary part of non-diagonal density matrix elements $\Im[\rho_{12}]=-\Im[\rho_{21}]$, with which we can effectively determine the imaginary part of the off-diagonal density matrix elements between exactly degenerate basis states, by using the products of velocities as physical constraints in the iterative QT procedure.

Unlike rotational and vibrational Quantum State Tomography, the coupling between electrons and nuclei severs as a strong system-bath interaction and the temporal evolution is not trivially dominated by the system Hamiltonian.
For electronic state we utilize Quantum Process Tomography (QPT), which is a systematic procedure to completely characterize a quantum process as a 'black box', by a sequence of measuring the inputs and outputs~\cite{Poyatos97:PRL78}. QPT has successfully retrieved quantum coherence dynamics in molecular systems, especially population and coherence transfer mechanism based on spectroscopic methods~\cite{Chuntonov13:JPCB117,YuenZhou11:JCP134}.

Consider a system of two electronic states coupled to the nuclear degrees of freedom. After laser excitation, the initial state $\ket{\Psi(0)}$ is prepared 
\begin{eqnarray}
\ket{\Psi(0)}=\ket{\psi_e(0)}\ket{\phi_e}+\ket{\psi_g(0)}\ket{\phi_g}
\,,
\end{eqnarray}
where $\phi_g,\phi_e$ are electronic ground state and excited state, and $\psi_g,\psi_e$ are corresponding nuclear wavepacket.  The electronic reduced density operator
\begin{eqnarray}
\hat{\rho}^{e}(t)=\int d\textbf{R} \BK{\textbf{R}}{\hat{\rho}^{tot}(t)}{\textbf{R}}=\sum_{a,b}\IP{\psi_a(t)}{\psi_b(t)}\ketBra{\phi_a}{\phi_b}
\,,
\end{eqnarray}
where the subscripts $a,b$ (and the following $c,d$) refer to the index of ground and excited states, and $\textbf{R}$ represents the nuclear degrees of freedom. Under the basis of electronic states, the temporal evolution of initial state can be expressed as a linear transformation~\cite{Sudars61:PR121}
\begin{eqnarray}\label{Eq:ProcMat}
\rho^{e}_{ab}(t) = \sum_{cd}\chi_{abcd}(t) \rho^{e}_{cd}(0)
\,.
\end{eqnarray}
The central object of QPT is to obtain the process matrix $\chi_{abcd}(t)$ by measuring the $\rho^{e}_{ab}(t)=\IP{\psi_a(t)}{\psi_b(t)}$ through ultrafast X-ray diffraction signal contributed from different initial state preparation $\rho^{e}_{ab}(0)$ determined by laser excitation parameters~\cite{Yuen11:PNAS43}. Together with the following properties of process matrix associated with trace preservation and Hermiticity~\cite{YuenZhou11:JCP134}
\begin{eqnarray}
\sum_a \chi_{aacd}(t)&=&\delta_{cd}\\
\chi_{abcd}(t)&=&\chi^*_{badc}(t)
\,,
\end{eqnarray}
the preparation of initial state $\rho^{e}_{ab}(0)$ should form a complete set so that the output state of any input can be predicted, and equivalently, the process matrix elements $\chi_{abcd}(t)$ can be solved from Eq.~\ref{Eq:ProcMat}. 

The temporal evolution of electronic state $\rho^{e}_{ab}(t)$ can be monitored by time-resolved X-ray diffraction. 
The gas phase off-resonance scattering signal is related to~\cite{Markus17:Strc4}
\begin{eqnarray}
I(\textbf{s},t)=\BK{\Psi(t)} {\hat{\sigma}^\dagger(\textbf{s},t)\hat{\sigma}(\textbf{s},t)}{\Psi(t)} =\sum_{a,b} \Bra{\psi_a(t)}\BK{\phi_a}{\hat{\sigma}^\dagger(\textbf{s},t)\hat{\sigma}(\textbf{s},t)}{\phi_b}\ket{\psi_b(t)}
\,,
\end{eqnarray}
where $\textbf{s}$ is the scattering momentum transfer defined in the main text and $\hat{\sigma}(\textbf{s},t)$ is the Fourier transform of electronic charge-density operator 
\begin{eqnarray}
\hat{\sigma}(\textbf{r})=\sum_{ab}\int d\textbf{r}_2\cdots d\textbf{r}_n \phi^*_a(\textbf{r},\textbf{r}_2,\cdots, \textbf{r}_n;\textbf{R})\phi_b(\textbf{r},\textbf{r}_2,\cdots, \textbf{r}_n;\textbf{R})
\,,
\end{eqnarray}
where $\textbf{r}_1,\cdots,\textbf{r}_n$ are the electron coordinates. The electronic density matrix $\rho^{e}_{ab}(t)=\IP{\psi_a(t)}{\psi_b(t)}$ can be retrieved by solving the Fredholm integral equation of the first kind, as is described in detail in the previous section (see Eq.~\ref{Eq:Fredh}). For each fixed time point, the integral can be written in the Riemann summation form
\begin{eqnarray}
I(\textbf{s}_i)=\sum_{ab}\sum_j \omega(\textbf{R}_j) \psi^*_a(\textbf{R}_j)\psi_b(\textbf{R}_j) \BK{\phi_a(\textbf{R}_j)}{\hat{\sigma}^\dagger(\textbf{s}_i;\textbf{R}_j)\hat{\sigma}(\textbf{s}_i;\textbf{R}_j)}{\phi_b(\textbf{R}_j)}
\,,
\end{eqnarray}
where $\omega(\textbf{R}_j)$ is the integration weight, $i=1,\cdots,M$ and $j=1,\cdots,N$ is the grid point index of $\textbf{s}$ and $\textbf{R}$, respectively.
The integral equation is converted to the matrix equation $I=\mathbf{K}\Pr$ by defining
\begin{eqnarray}
I&=&{ \left( \begin{array}{c}
I(\textbf{s}_1)\\\vdots\\I(\textbf{s}_M)\end{array} \right )}\\
\mathbf{K}&=&{
\left( \begin{array}{ccc}
\BK{\phi_a(\textbf{R}_1)}{\hat{\sigma}^\dagger(\textbf{s}_1;\textbf{R}_1)\hat{\sigma}(\textbf{s}_1;\textbf{R}_1)}{\phi_b(\textbf{R}_1)}& \cdots&
\BK{\phi_a(\textbf{R}_N)}{\hat{\sigma}^\dagger(\textbf{s}_1;\textbf{R}_N)\hat{\sigma}(\textbf{s}_1;\textbf{R}_N)}{\phi_b(\textbf{R}_N)}\\
\vdots& \ddots & \vdots\\
\BK{\phi_a(\textbf{R}_1)}{\hat{\sigma}^\dagger(\textbf{s}_M;\textbf{R}_1)\hat{\sigma}(\textbf{s}_M;\textbf{R}_1)}{\phi_b(\textbf{R}_1)}& \cdots&
\BK{\phi_a(\textbf{R}_N)}{\hat{\sigma}^\dagger(\textbf{s}_M;\textbf{R}_N)\hat{\sigma}(\textbf{s}_M;\textbf{R}_N)}{\phi_b(\textbf{R}_N)}
\end{array}
\right )}\\
\Pr&=&{ \left( \begin{array}{c}
\psi^*_a(\textbf{R}_1)\psi_b(\textbf{R}_1)\omega(\textbf{R}_1)\\\vdots\\\psi^*_a(\textbf{R}_N)\psi_b(\textbf{R}_N)\omega(\textbf{R}_N)
\end{array} \right )}
\,,
\end{eqnarray}
The subscripts $a,b$ occurred in $\mathbf{K}$ and $\Pr$ need to traverse the ground state and excited state. Thus, in principle, after solving the matrix equation for $\Pr$, we simultaneously recovered the nuclear state $\psi_a(\textbf{R})$ and electronic state
\begin{eqnarray}
\rho^{e}_{ab}=\sum_j\omega(\textbf{R}_j)\psi^*_a(\textbf{R}_j)\psi_b(\textbf{R}_j)
\end{eqnarray}
recorded by ultrafast diffraction. However, the algorithm will be strongly restricted by the dimension of nuclear configuration, which is usually much larger than the 2D diffraction pattern.

Much simplification can be made if we only focus on the electronic density matrix. For most cases the nuclear wavepacket moves around the equilibrium point, and the electronic wavefunction can be approximated to fixed reference nuclear configuration. 
If we consider the diabatic representation description where electronic wavefunction $\phi_a$ and electronic charge-density operator $\hat{\sigma}$ do not change with nuclear geometry
\begin{eqnarray}
I(\textbf{s},t)=\sum_{a,b} \IP{\psi_a(t)}{\psi_b(t)}\BK{\phi_a}{\hat{\sigma}^\dagger(\textbf{s},t)\hat{\sigma}(\textbf{s},t)}{\phi_b}=\sum_{a,b}\rho^{e}_{ab}(t)\BK{\phi_a}{\hat{\sigma}^\dagger(\textbf{s},t)\hat{\sigma}(\textbf{s},t)}{\phi_b}
\,.
\end{eqnarray}
By choosing a suitable reference nuclear configuration, the temporal evolution of $\rho^{e}_{ab}(t)$ can be solved directly.

Throughout the paper, we focus on recovering the density matrix, which is interconnected with the Wigner function $W(q,p)$ via the overlapping formula,
\begin{eqnarray}
\label{Eq:Wigner_DM}
\rho_{mn}&=&\mathrm{Tr}[\hat{\rho}\ket{n}\Bra{m}]\nonumber\\
&=&\frac{1}{2\pi}\int_{-\infty}^{\infty}\,dq\int_{-\infty}^{\infty}\,dp W(q,p)W_{\ketBra{n}{m}}(q,p)
\,,
\end{eqnarray}
where $W_{\hat{\mathcal{O}}}(q,p)=({1}/{2\pi})\int dx \exp(-ipx)\BK{q-\frac{x}{2}}{\hat{\mathcal{O}}}{q+\frac{x}{2}}\,.$
Especially, the Wigner function can be expressed in terms of the density operator $\hat{\rho}$ as $W(q,p)=W_{\hat{\rho}}(q,p)$.

\bibliographystyle{naturemag}
\bibliography{QT}